\documentclass[prl,twocolumn,superscriptaddress,showpacs,floatfix]{revtex4-1}
\usepackage{epsfig,amsmath,amssymb,latexsym}

\usepackage{subfigure}
\usepackage{xcolor}
\usepackage{soul}
\usepackage{ulem}

\def\afs{$A_{0.8}$Fe$_{1.6}$Se$_2$}
\def\kfs122{K$_y$Fe$_2$Se$_2$}
\def\kfsx22{K$_{1-x}$Fe$_2$Se$_2$}

\def\bfa122{BaFe$_2$As$_2$}
\def\fs11{FeSe}
\def\kfsxy{K$_y$Fe$_{2-x}$Se$_2$}
\def\a245{$A_2$Fe$_4$Se$_5$}
\def\neelfm{N\'{e}el-AM}
\def\rb245{Rb$_2$Fe$_4$Se$_5$}
\def\k245{K$_2$Fe$_4$Se$_5$}
\def\tl245{Tl$_2$Fe$_4$Se$_5$}
\def\cs245{Cs$_2$Fe$_4$Se$_5$}
\def\na245{Na$_2$Fe$_4$Se$_5$}

\begin{document}

\title{Pressure Induced Altermagnetism in Layered Ternary Iron-Selenides} \preprint{1}

\author{Zilong Li}
\author{Xin Ma}
 \affiliation{Center for Correlated Matter and School of Physics, Zhejiang University, Hangzhou 310058, China}

\author{Siqi Wu}
 \affiliation{Department of Physics, The Hong Kong University of Science and Technology, Clear Water Bay, Kowloon, Hong Kong, China}

\author{H.-Q. Yuan}
 \affiliation{Center for Correlated Matter and School of Physics, Zhejiang University, Hangzhou 310058, China}
 \affiliation{Institute for Advanced Study in Physics, Zhejiang University, Hangzhou 310058, China}

\author{Jianhui Dai}
\email[E-mail address: ]{daijh@hznu.edu.cn}
 \affiliation{School of Physics, Hangzhou Normal University, Hangzhou 310036, China}
 \affiliation{Institute for Advanced Study in Physics, Zhejiang University, Hangzhou 310058, China}

\author{Chao Cao}
 \email[E-mail address: ]{ccao@zju.edu.cn}
 \affiliation{Center for Correlated Matter and School of Physics, Zhejiang University, Hangzhou 310058, China}
 \affiliation{Institute for Advanced Study in Physics, Zhejiang University, Hangzhou 310058, China}

\date{Feb. 21, 2025}

\begin{abstract}
Employing first-principles based calculations, we reexamined the high-pressure phases of the vacancy-ordered iron-selenides, i.e. \a245\ phase. A magnetic transition from the block-spin antiferromagnetic phase to \neelfm\ phase is observed under high pressure when the iron-vacancy order is preserved. The transition is first-order, driven by the collapse of $c$-axis and accompanied by an insulator-metal transition. In addition, the \neelfm\ phase is a rare example of {\it intrinsic} altermagnetism that does not require ligand atoms nor supercell construction to break the spatial inversion or translational symmetry between the two spin sublattices, with a spin-splitting of the band structure as large as 300 meV. If the re-entrant superconducting phase of \kfsxy\ under high pressure emerges from the \neelfm\ normal state, an equal-spin triplet pairing would be naturally favored.
\end{abstract}

\maketitle


{\it Introduction.} The concept of altermagnetism has attracted considerable attention recently\cite{PhysRevX.12.040501}. Theoretically, altermagnetism is closely related to and resembles anisotropic Fermi-liquid instabilities\cite{PhysRevLett.93.036403,PhysRevB.75.115103}, and is characterized by spin-orbit coupling (SOC) free compensated spin-up and spin-down sublattices that are spatially connected by rotational symmetry\cite{PhysRevX.12.031042,PhysRevX.14.031038}, instead of the translational or inversion symmetry in conventional antiferromagnetism (AFM). Therefore, the spin-degeneracy is lifted in the altermagnetism at general $\mathbf{k}$-points, since the $\lbrace\mathcal{T}||\mathbf{t}\rbrace$ or $\lbrace\mathcal{T}||i\rbrace$ symmetry is broken\cite{PhysRevX.12.031042,PhysRevX.14.031038,Li:2022aa,Krempasky2024-dj} ($\mathcal{T}$, $\mathbf{t}$ and $i$ represent time-reversal, lattice translation and inversion symmetry, respectively). Such feature is desirable in spintronics\cite{PhysRevLett.129.276601,PhysRevX.12.040501}, and may lead to exotic superconducting pairing states\cite{mazin2022notes,PhysRevX.12.040501,Zhang:2024aa,PhysRevB.109.134515,PhysRevB.109.L201404,PhysRevB.110.L060508,PhysRevB.111.L220403}. However, material candidates for altermagnetism-related superconductivity (SC) is currently still lacking and remain elusive.

\begin{figure}[h]
  \includegraphics[width=8cm]{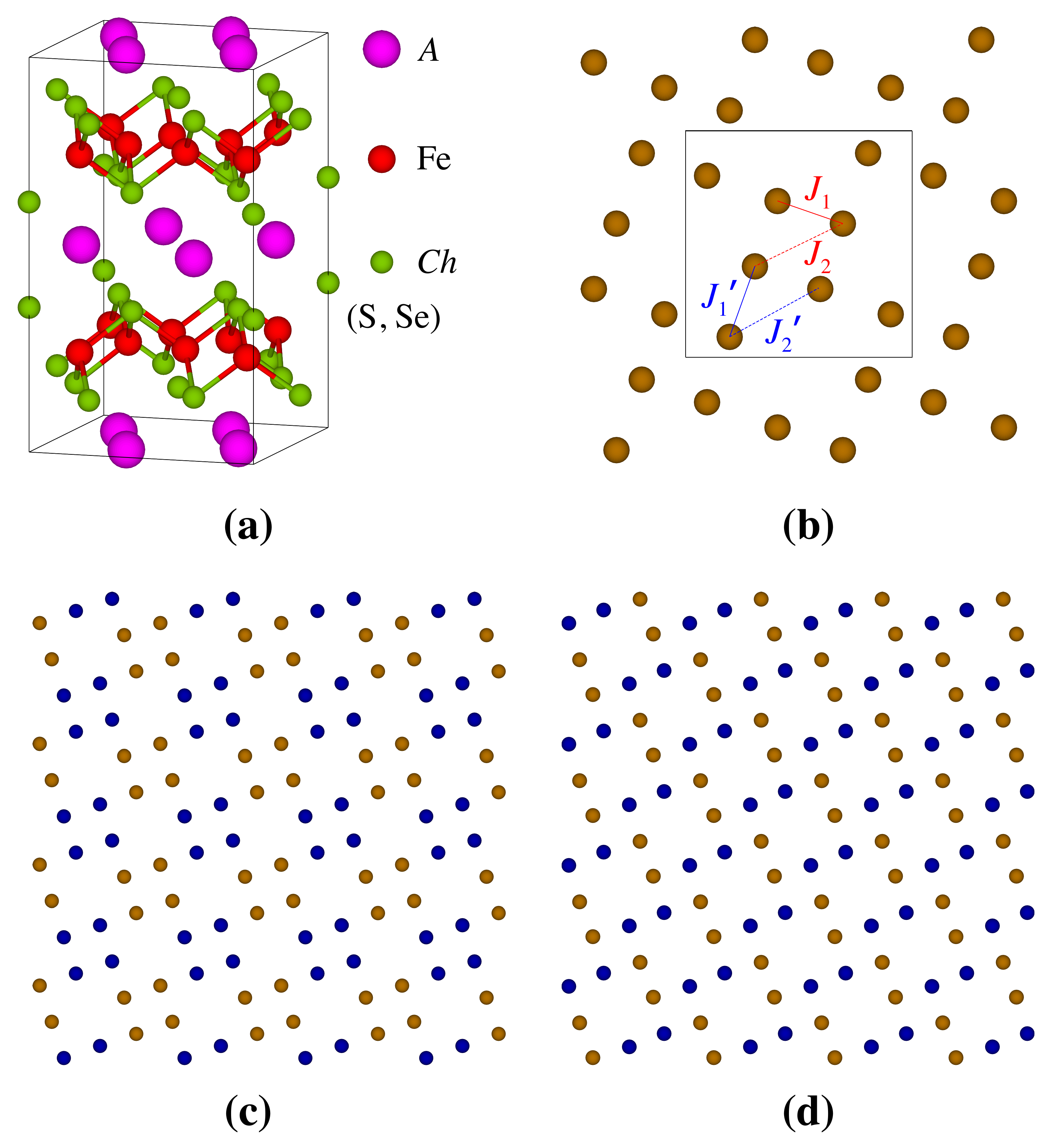}
  \caption{Crystal structure and dominant magnetic phases. (a) 3D crystal structure showing the anti-PbO-type FeSe layer. (b) The Fe-plane and definition of exchange interactions. The black square shows the in-plane unit cell. (c) BS-AFM phase. (d) \neelfm\ phase. In (b-d), only Fe atoms are shown; blue/brown atoms represent Fe atoms with up/down moments, respectively. \label{fig:struct}}
\end{figure}

Of the iron-based superconductors, the ternary iron-selenide compounds $A_y$Fe$_{2-x}$Se$_2$ ($A$=Tl, K, Rb, Cs) are of particular interest\cite{PhysRevB.82.180520,Wang_2011,Fang_2011}. Structurally, these compounds are similar to the 122-type iron-pnictides\cite{PhysRevLett.101.107007}. The anti-PbO-type FeSe layers are intercalated by $A$ atoms, leaving certain amount of Fe-vacancies in the Fe-square lattice(FIG. \ref{fig:struct}). At $x=0.4$, a robust $\sqrt{5}\times\sqrt{5}$ vacancy-ordered superstructure forms, and the ground state becomes block-spin AFM (BS-AFM)\cite{Bao_2011,PhysRevLett.107.056401,PhysRevB.83.233205,PhysRevLett.107.137003} (FIG. \ref{fig:struct}c). In addition, the compound is insulating at $y=0.8$\cite{PhysRevLett.107.056401,PhysRevB.83.233205}. Therefore, the \a245(\afs) distinguishes itself from rest of the family and is often referred to as the 245-phase. The ordered Fe-vacancy is found to be quite robust and universally exists not only in $A_y$Fe$_{2-x}$Se$_2$ and $A_y$Fe$_{2-x}$S$_2$\cite{Toulemonde_2013,Bao_2015,PhysRevB.100.094108,PhysRevB.101.054516,Chen:2021aa}, but also in $\beta$-Fe$_{1-x}$Se\cite{pnas.1321160111,PhysRevB.95.174523,fphy.2020.567054}, even under very high pressure up to $\sim$16 GPa\cite{Sun:2012aa,PhysRevB.85.214519}. In addition, a re-emerging superconductiviting phase between 10.5 GPa and 13.2 GPa was observed experimentally\cite{Sun:2012aa}. However, the superconducting phase of $A_y$Fe$_{2-x}$Se$_2$ was the focus of previous studies, and much less attention was paid to the insulating 245-phase. The stability of such vacancy order and its influence on the magnetic as well as the insulator-metal transitions in this material family deserve more investigations. 

\begin{figure*}
  \includegraphics[width=16cm]{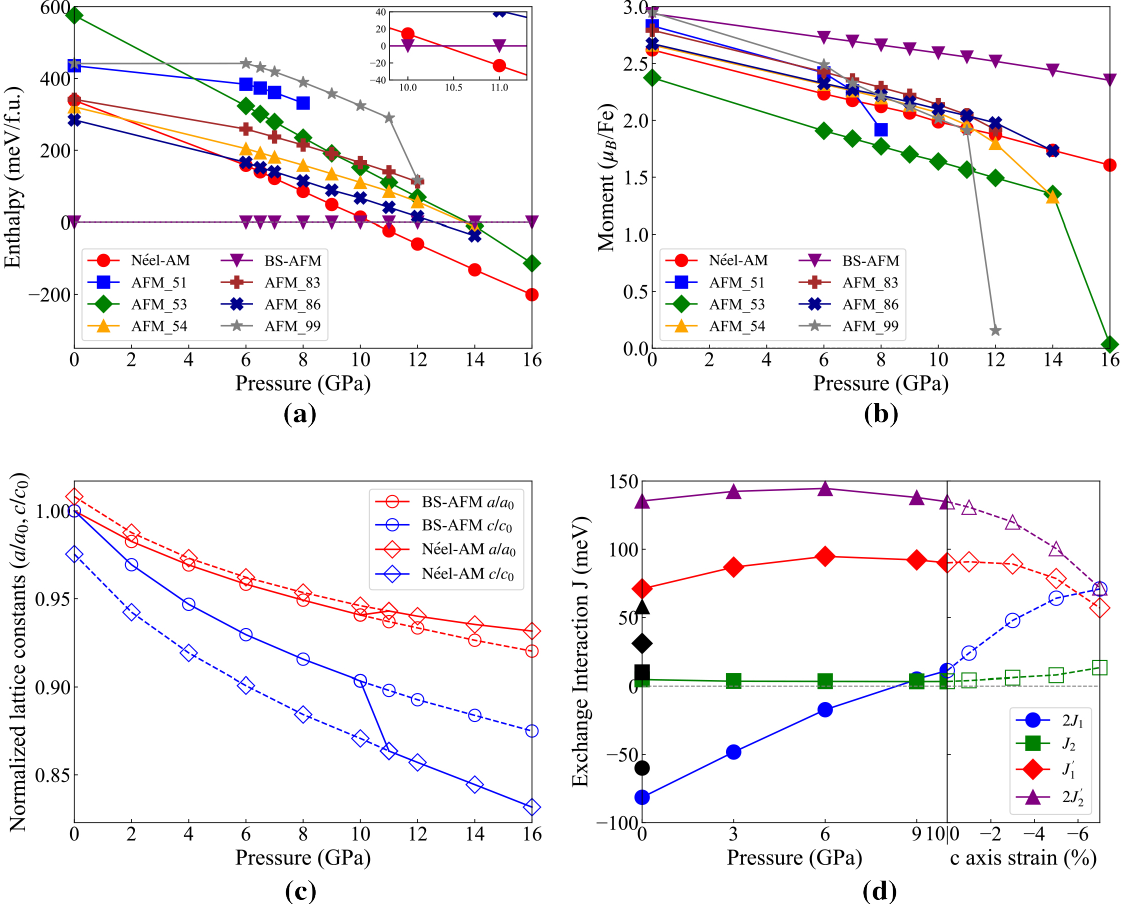}
  \caption{(a) Enthalpy of different \rb245\ phases under pressure. All energies are relative to the BS-AFM phase at the same pressure. (b) Ordered moments of different \rb245\ phases under pressure. For the AFM-51 phases, the configuration cannot be stabilized after 8 GPa; for the AFM-54 and AFM-86 configurations, the ordered moment becomes negligibly small after 14 GPa. (c) The normalized lattice constants $a/a_0$ and $c/c_0$. Both $a_0$ and $c_0$ refers to the values of the ground state (BS-AFM) at ambient pressure. The solid line follows the lowest energy state. (d) The fitted exchange parameters $J$ in the extended $J_1-J_2$ model. Left panel shows the pressure dependence, whereas the right panel shows the $c$-stress dependence. The black symbols at 0 GPa are experimental results from Ref. \cite{PhysRevB.87.100501}.\label{fig:enmom}}
\end{figure*}
In this paper, we reexamine the high-pressure phase of the vacancy-ordered 245-phase selenides, i.e. \a245. We found a first-order transition from the BS-AFM phase to the \neelfm\ phase in all these compounds in addition to $A=$(Tl, K) we found earlier\cite{cao2011blockspinmagneticphase}, under certain hydraulitic pressure $p_c$. We address the issue of overestimated magnetic interactions using DFT+DMFT calculations\cite{method:dmft2,method:ctqmc_dmft}, and found the competing stripe-like AFM phase is not favored. The magnetic phase transition is driven by the collapse of $c$-axis, and is accompanied by an insulator-metal transition. In addition, the \neelfm\ phase is altermagnetic, exhibiting spin-splitted band structure up to 300 meV. Unlike RuO$_2$\cite{Smejkal:aa,Shao:2021aa} or other proposed altermagnetic compounds\cite{PhysRevX.12.031042}, the \neelfm\ phase does not require ligand atoms to break the inversion or translation symmetry between the spin-sublattices, and its magnetic unit cell is the same size as its nonmagnetic primitive cell. As a result, the spin-splitting in the \neelfm\ phase is more robust since it does not depend on the ligand atom positions that are susceptible to defect, external pressure, temperature, or doping. Therefore, the \neelfm\ pattern is an example of {\it intrinsic} altermagnetism. This renders the high-pressure superconducting phase of \kfsxy\ between 10.5 GPa and 13.2 GPa a possible candidate for studying inter-play between superconductivity and altermagnets.

{\it Magnetic Phase Transition under Pressure.} Employing a high-throughput method\cite{Xu:2025aa}, we have searched the ground state magnetic phase of \a245 under different pressure. There are 8 inequivalent AFM configurations for the primitive cell (8 Fe atoms) (see SI section B and FIG. S-2 for complete list) and 508 inequivalent AFM configurations for size-2 supercell (16 Fe atoms). Previously, an extended $J_1-J_2$ model has been identified to well explain the magnetic configurations\cite{PhysRevLett.107.056401,PhysRevB.86.020402,PhysRevB.87.100501}, suggesting that the magnetic interactions are fairly local in the 245 phase. In this model, the effective spin Hamiltonian can be written as: $H=J_1/2\sum_{\langle i,j\rangle}\mathbf{S}_i\cdot \mathbf{S}_j+J'_1/2\sum_{\lbrace i,j\rbrace}\mathbf{S}_i\cdot \mathbf{S}_j+J_2/2\sum_{\langle\langle i,j\rangle\rangle}\mathbf{S}_i\cdot \mathbf{S}_j+J'_2/2\sum_{\lbrace\lbrace i,j\rbrace\rbrace}\mathbf{S}_i\cdot \mathbf{S}_j$, where $\langle i,j\rangle$ and $\lbrace i,j\rbrace$ denote intra-block and inter-block nearest neighbors, $\langle\langle i,j\rangle\rangle$ and $\lbrace\lbrace i,j\rbrace\rbrace$ are intra-block and inter-block next nearest neighbors, respectively. Therefore, we constrain the search within the primitive cell in most cases.  In addition, neutron scattering experiment also reveals small inter-layer interactions ($J_cS^2<$1.0 meV)\cite{PhysRevB.87.100501}, thus we consider in-plane configurations only. The stability of the \neelfm\ ground state is further confirmed based on extended $J_1-J_2$ parameters at 12 GPa using Monte Carlo simulations(see SI section C and FIG. S-3 for details). 

The BS-AFM phase is the ground state for all the compounds under ambient pressure, and is energetically overwhelmingly favored. For example, the second lowest energy phase is 280 meV/f.u. higher than the BS-AFM phase for \rb245, and the \neelfm\ phase is 340 meV/f.u. higher than the BS-AFM phase (FIG. \ref{fig:enmom}a). At 10 GPa, the BS-AFM remains as the ground state, while the \neelfm\ phase becomes second lowest, and its enthalpy $H=E+pV$ is only 17 meV/f.u. higher. At 11 GPa and beyond (to at least 16 GPa), the \neelfm\ phase becomes the lowest enthalpy state. Between 12 GPa and 16 GPa, its enthalpy is at least 30 meV/f.u. lower than the second lowest state. While a stripe-like AFM was shown to be a low-energy phase and compete with the \neelfm\ phase at the DFT level\cite{cao2011blockspinmagneticphase}, the magnetism of the stripe-like AFM phase is known to be severely overestimated in normal DFT calculations, and the calculated staggered moment is several times larger than experimental value\cite{PhysRevB.78.085104}. This is in sharp contrast to the BS-AFM phase whose magnetic moments from DFT calculations are in good agreement with experiment\cite{Bao_2011,PhysRevLett.107.056401}. In addition, our DFT+DMFT calculations at 12 GPa show that the enthalpy of the stripe-like AFM phase is in fact $\sim$100 meV/f.u. higher than the \neelfm\ phase, and the ordered moments are consistent with experimental observation (see SI section E for details). We also confirm that no AFM configuration for size-2 supercell has lower enthalpy than \neelfm\ phase at 12 GPa. Thus, we conclude that the ground state magnetic phase is \neelfm\ between 11 and 16 GPa in \rb245. Similarly, this phase transition can be identified around 11 GPa in \k245, 13 GPa in \na245, 12 GPa in \tl245, and 9 GPa in \cs245, respectively (see SI section F and FIG S-7 for details). 

\begin{figure*}
  \includegraphics[width=16cm]{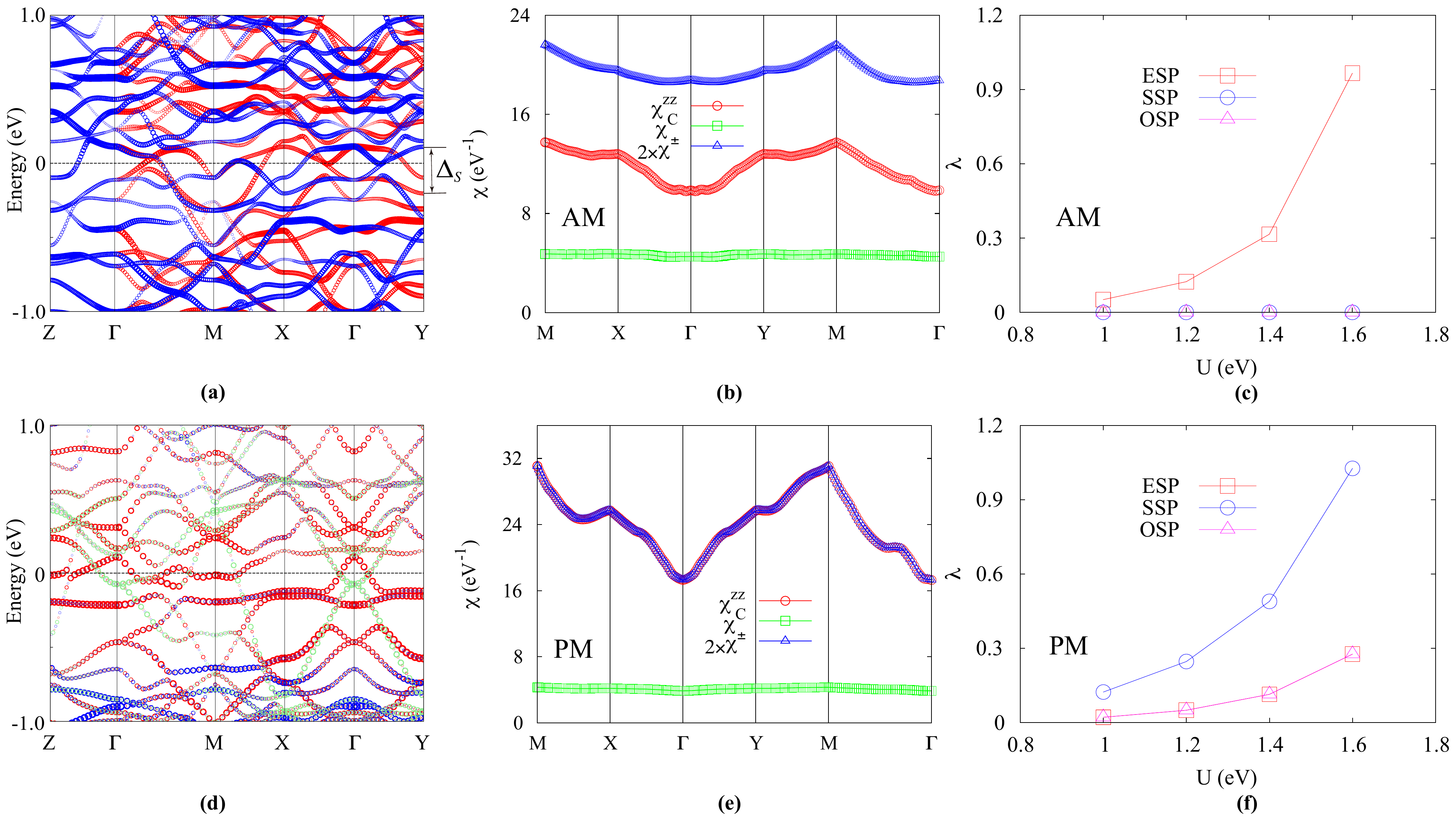}
  \caption{(a,d) Band structure, (b,e) charge ($\chi^C$) and longitudinal/transverse spin ($\chi^{zz}$/$\chi^{\pm}$) susceptibility, and (c,f) pairing strength $\lambda$ of leading channels of Rb$_2$Fe$_4$Se$_5$ at 14 GPa. Upper/lower panels are based on \neelfm\ ordered/paramagnetic normal state, respectively. In panel (a), the red/blue color represents states of opposite spin, and the sizes of the circles are proportional to the weight of Fe-3$d_{zx(y)}$ orbitals. $\Delta_s$ indicates the spin-splitting at X or Y. In panel (d), the sizes of the red, blue and light-green circles are proportional to the weight of Fe-3$d_{zx(y)}$, Fe-3$d_{z^2}$ and Fe-3$d_{x^2-y^2}$ orbitals, respectively.\label{fig:bsdos}}
\end{figure*}


External pressure effect is also evident on the magnetic moment. Under ambient pressure, \rb245\ exhibits 2.9 $\mu_{\mathrm{B}}$/Fe ordered moment (FIG. \ref{fig:enmom}b), in good agreement with experimental observation of 3.31 $\mu_{\mathrm{B}}$/Fe \cite{Bao_2011}. It is almost linearly suppressed to 2.56 $\mu_{\mathrm{B}}$/Fe at 10 GPa before the transition, then abruptly reduced to 1.91 $\mu_{\mathrm{B}}$/Fe in the \neelfm\ phase at 11 GPa. Further increasing the pressure, the ordered moment is again linearly suppressed to 1.55 $\mu_{\mathrm{B}}$/Fe at 16 GPa. We also note that the AFM-51, AFM-54, AFM-99, AFM-83 and AFM-86 phases become unstable between 8 and 14 GPa, and AFM-53 becomes unstable at 16 GPa, as their ordered moments are quickly suppressed to less than 0.1 $\mu_{\mathrm{B}}$/Fe and the total enthalpies approach to NM phases. Therefore, these phases become nonmagnetic at high pressure in calculation. The ordered moment is robust in both the BS-AFM and the \neelfm\ phase, as no sign of sudden drop can be observed up to 16 GPa. The result also supports a competition only between the BS-AFM phase and the \neelfm\ phase under high pressure.

{\it Evolution of Crystal Structure.} The space group of fully relaxed BS-AFM phase \rb245\ at ambient pressure is $I$4/m (No. 87). Each primitive cell contains 2 chemical formula units, and all 8 Fe-sites are equivalent. The lattice constants $a$ and $c$ of \rb245\ are $a_0=$8.730\AA\ and $c_0=$14.888\AA\ under ambient pressure in the BS-AFM phase, respectively. These values are in good agreement with the experimental observation\cite{PhysRevLett.107.137003}, with less than 2\% error. At 8 GPa, $a$ and $c$ are reduced by 5.0\% and 8.9\% compared to $a_0$ and $c_0$, respectively, which is also in perfect agreement with the experimental observation\cite{Ye_2014}. From 10 GPa to 11 GPa across the magnetic phase transition, the space group remains unchanged. However, the lattice constant $c$ is suddenly reduced by 4.4\% while the change in $a$ is less than 0.3\%. Thus, the structural transition is signaled by the collapse of $c$-axis. The $c$-collapse effect under pressure was indeed reported in K$_y$Fe$_{2-x}$S$_2$\cite{JPSJ.86.033705}, and can be inferred from the discontinuity in $d_{\mathrm{Fe-Se}}$ in \kfsxy\cite{PhysRevB.88.180506}. Although such transition was attributed to formation of disordered vacancy phase previously\cite{PhysRevB.88.180506,Yamamoto:2016aa}, the transition from BS-AFM to \neelfm\ phase provides an alternative explanation to the observed effect. Since both the deviation from stoichemical composition and thermal fluctuations enhance disorder effect\cite{Bao_2015}, a careful characterization of stoichemical semiconducting sample under pressure may resolve this issue. Furthermore, the dynamical stability of the \neelfm\ phase is also verified by calculating its phonon spectrum at 12 GPa (see SI section C and FIG. S-4 for details). 

{\it Spin Model Analysis.} We briefly discuss the formation of the \neelfm\ phase under the framework of the spin models. For the vacancy-ordered 245 phase, the simplest spin model would be $J_1-J_1'$ model that considers nearest neighboring interactions only but differentiates the intra-block ($J_1$) and inter-block ($J_1'$) interactions (FIG. \ref{fig:struct}). In such case, it is apparent that 4 possible magnetic phases exist depending on the signs of $J_1$ and $J_1'$: checker-board AFM ($J_1>0$ and $J_1'>0$), BS-AFM ($J_1<0$ and $J_1'>0$), \neelfm\ ($J_1>0$ and $J_1'<0$), and FM ($J_1<0$ and $J_1'<0$). Thus, both BS-AFM and \neelfm\ phase are dominating phases depending on the overall nature of intra-block and inter-block spin-interactions. If the next-nearest neighboring interactions are included, i.e. the extended $J_1-J_2$ model, the above statement remains valid as shown by Yu {\it et al.} using the classical Monte-Carlo simulations\cite{PhysRevB.84.094451}. Using the classical expressions for different phases\cite{PhysRevB.84.094451,cao2011blockspinmagneticphase}, the \neelfm\ phase is energetically favored over the BS-AFM, checkerboard AFM and the FM phase, if $2J_1>J_1'$, $2J_2'>J1'$, and $J_1+J_2'>0$, respectively. In addition, the energy of the \neelfm\ phase is lower than the stripe-like AFM if $3J_1+J_2'>J_1'+J_2$. 

We fitted the exchange interactions $J$s of the extended $J_1$-$J_2$ model under different pressure (see SI section G and TAB. S-II for details). At ambient pressure, the fitted $J_1=-40.7$ meV/$S^2$, $J_1'=71.0$ meV/$S^2$, $J_2=4.9$ meV/$S^2$ and $J_2'=67.7$ meV/$S^2$. These are in reasonable agreement with previous experiment\cite{PhysRevB.87.100501} (FIG. \ref{fig:enmom}d), although both inter-block interactions $J_1'$ and $J_2'$ seem to be substantially overestimated. Under pressure, $J_2$ is nearly unchanged, while both $J_1'$ and $J_2'$ slightly enhances below 6 GPa and gradually decreases. In contrast, the pressure drastically alters $J_1$, which changes its sign close to 9 GPa, where all interactions become antiferromagnetic. Nevertheless, even at 10 GPa, $J_1$ is too small compared to $J_1'$ to drive the magnetic phase transition. Thus, we believe that the phase transition is first-order, induced by the sudden reduction of $c$. To confirm this, we have also performed calculations with reduced $c$ with fixed $a$ at 10 GPa. Both $J_1'$ and $J_2'$ are drastically reduced if $c$ is under stress, while $J_1$ quickly further enhances. The condition $2J_1>J_1'$ is fulfilled when the $\Delta c/c$ reaches -7\%. It is therefore reasonable to postulate that a uniaxial pressure in $c$-direction is more effective to tune the exchange interactions and the \neelfm\ phase may be achieved at lower pressure in such case.


{\it Evolution of Electronic Structure.} Under ambient pressure, \rb245\ is an insulator with an indirect gap $E_g\approx 580$ meV, formed by the Fe-3$d_{zx(y)}$ states at X ($\pi$, 0, 0) and the Fe-3$d_{z^2}$ state at $\Gamma$ (FIG. S-8a). Under pressure, the band gap $E_g$ is monotonically suppressed to 290 meV at 10 GPa (FIG. S-8b). In contrast, the \neelfm\ phase is always metallic, therefore the magnetic phase transition is accompanied by a insulator-metal transition simultaneously. 

As shown in Fig. \ref{fig:struct}(d), the spin-up and spin-down sublattices of the \neelfm\ phase are spatially connected by the $C_4$ rotation, thus it is altermagnetic. Similar to the ``failed AF-SOD" phase in CsCr$_3$Sb$_5$\cite{Xu:2025aa}, the \neelfm\ pattern does not require ligand atoms to break the inversion or translation symmetry between these two sublattices. However, unlike the ``failed AF-SOD" phase, the magnetic unit cell of the \neelfm\ phase is the same size as the chemical unit cell. Therefore, the \neelfm\ phase is also different from the supercell altermagnets\cite{PhysRevB.109.094425}, and we denote it as {\it intrinsic} altermagnetism. As shown in FIG. \ref{fig:bsdos}(a), the spin degeneracy is lifed at general $\mathbf{k}$-points in \neelfm\ phase \rb245, although the spatial inversion symmetry is preserved in the compound. Therefore, $\epsilon_{\mathbf{k}\uparrow}=\epsilon_{\mathbf{-k}\uparrow}\neq \epsilon_{\mathbf{k}\downarrow}=\epsilon_{\mathbf{-k}\downarrow}$. The spin-splitting $\Delta_s$ close to $E_F$ at X ($\pi$, 0, 0) or Y (0, $\pi$, 0) can be estimated by the Fe-3$d_{zx(y)}$ states to be $\sim$ 300 meV at 14 GPa. Finally, in the \neelfm\ phase, the electronic states close to the Fermi level are dominated by Fe-3$d_{zx(y)}$ and Fe-3$d_{x^2-y^2}$ orbitals. 

{\it Discussions.} Our prediction of the \neelfm\ phase is highly dependent on the formation of the $\sqrt{5}\times\sqrt{5}$ vacancy-order. However, controversial results regarding whether the vacancy-order can survive under high pressure have been reported\cite{ic201160y,Sun:2012aa,PhysRevB.85.214519,Ye_2014,JPSJ.86.033705}. We note that these experiments were performed with superconducting samples, whose composition deviates from the perfect 245-phase. In addition, the iron vacancy-ordered phase is much more favored at low temperatures\cite{ic201160y,Ye_2014}. Nevertheless, the synchrotron results of 245 selenides are in agreement that the $I4/m$ structure persists to at least 12 GPa\cite{ic201160y,PhysRevB.85.214519}. Therefore, it is possible to realize the \neelfm\ phase under sufficiently high pressure in 245 selenides. In fact, previous high pressure neutron scattering experiment has observed disappearance of the magnetic Bragg peak at lower pressure than the vacancy-order Bragg peak\cite{Ye_2014}. We note here that the propagation wave vector of \neelfm\ phase is $\mathbf{Q}=0$\citep{cao2011blockspinmagneticphase}, and always coincide with the structural peaks. Therefore, the disappearance of the magnetic peaks does not preclude the existence of the \neelfm\ phase.

It is also worth noting that a re-emerging superconducting phase between 10.5 GPa and 13.2 GPa was observed experimentally\cite{Sun:2012aa}. Given the intrinsic phase separation in the superconducting samples, it is possible that the vacancy order survives and the \neelfm\ order becomes present under this pressure. In such case, there are at least three possible scenarios relevant to the SC. In the first scenario, the superconducting phase is different from the \neelfm\ phase. Then the system becomes natural realization of a superconductor in close proximity of altermagnetism, where finite-momentum Cooper pairing may occur\cite{Zhang:2024aa,PhysRevB.110.L060508}. It was also proposed that finite-energy pairing in this case may lead to closed superconducting gap and a pair of mirage gaps\cite{PhysRevB.109.L201404}. In the second scenario, the SC emerges from a paramagnetic normal state by suppressing the \neelfm\ order; while in the last scenario, the SC emerges directly from a \neelfm\ order normal state. 


We shall elaborate slightly further on the second and the third cases here. In the second scenario, the time reversal symmetry is present in the normal state, and the pairing is likely to be mediated by residual spin fluctuation from suppressed long range order. Employing random phase approximations (RPA), the calculated spin susceptibility exhibits significant enhancement at zone corners when on-site interactions on Fe atoms are employed, and the relation $\chi^{zz}=2\chi^{\pm}$ holds due to the SU(2) symmetry [FIG. \ref{fig:bsdos}(e), details to the RPA calculations can be found in SI section I]. Meanwhile, the charge susceptibility $\chi^C$ is suppressed and featureless. Thus the leading spin fluctuation is antiferromagnetic, favoring a spin-singlet pairing [FIG. \ref{fig:bsdos}(f)]. In the third scenario, since the SOC is negligible, the SC gap functions can still be organized by spin characters, giving rise to a singlet channel (SSP) $\Delta^0=(\Delta_{\uparrow\downarrow}-\Delta_{\downarrow\uparrow})/\sqrt{2}$, triplet opposite spin channel (OSP) $\Delta^z=(\Delta_{\uparrow\downarrow}+\Delta_{\downarrow\uparrow})/\sqrt{2}$, and 2 triplet equal spin channels (ESP) $\Delta_{\uparrow\uparrow}$ and $\Delta_{\downarrow\downarrow}$. Since the spatial inversion symmetry is preserved, $\epsilon_{\mathbf{k}\sigma}=\epsilon_{\mathbf{-k}\sigma}$ ($\sigma=\uparrow$,$\downarrow$) is satisfied in the \neelfm\ phase. Therefore, both ESP channels remain active. In contrast, the spin-splitting $\Delta_s=|\epsilon_{\mathbf{k}\uparrow}-\epsilon_{\mathbf{k}\downarrow}|$ close to $E_F$ is order of 100 meV except a few degenerate nodal lines, thus neither SSP nor OSP channel is favored. We have solved the linearized gap equations in this case assuming momentum dependent pairing potential due to spin fluctuation as well. The charge susceptibility $\chi^C$ remains suppressed, but the transverse spin susceptibility $\chi^{\pm}$ is much more enhanced than the longitudinal spin susceptibility $\chi^{zz}$ in this case [FIG. \ref{fig:bsdos}(b)]. The pairing strengths of the leading SSP or OSP channels are negligibly small compared to those of ESP channels ($\mathrm{max}\lbrace \lambda^{\mathrm{SSP}}, \lambda^{\mathrm{OSP}}\rbrace/\mathrm{max}\lbrace \lambda^{\mathrm{ESP}} \rbrace<10^{-5}$) [FIG. \ref{fig:bsdos}(c)], thus a triplet ESP dominates in this case. Previously, a similar case assuming electron-phonon coupling was also studied\cite{leraand2025phononmediatedspinpolarizedsuperconductivityaltermagnets}. It was found that electron-phonon coupling can also lead to spin-triplet pairing if the interaction is sufficiently anisotropic.

In conclusion, we have performed systematic first-principles calculations to study the electronic structure and magnetism of \a245\ compounds under high pressure. We argue a possible magnetic-phase transition from the BS-AFM to \neelfm\ phase, accompanied by an insulator-metal transition at high pressure before the ordered iron-vacancy structure collapse. The \neelfm\ phase is an {\it intrinsic} altermagnetism that does not require ligand atoms nor supercell construction to break the inversion or translational symmetry between the spin up/down sublattices. It leads to large spin-splitted band structure as well as spin-polarized Fermi surfaces. Finally, if the re-emerging superconducting phase in \kfsxy\ under high pressure emerges from the \neelfm\ normal phase, an ESP triplet pairing is naturally expected.

\begin{acknowledgments}
We would like to thank Lunhui Hu, Hua Chen, Chenchao Xu, Minghu Fang, Hangdong Wang, Yuanfeng Xu, Ming Shi, Yang Liu and Yu Song for the stimulating discussions. This work has been supported by the National Key R\&D Program of China (Nos. 2024YFA1408303 \& 2022YFA1402202) and the National Natural Science Foundation of China (Nos. 12274364 \& 12274109). The calculations were performed on the Quantum Many-Body Computing Cluster at Zhejiang University and High Performance Computing Center of Hangzhou Normal University.
\end{acknowledgments}

 \bibliography{245-p}

\begin{thebibliography}{49}%
\makeatletter
\providecommand \@ifxundefined [1]{%
 \@ifx{#1\undefined}
}%
\providecommand \@ifnum [1]{%
 \ifnum #1\expandafter \@firstoftwo
 \else \expandafter \@secondoftwo
 \fi
}%
\providecommand \@ifx [1]{%
 \ifx #1\expandafter \@firstoftwo
 \else \expandafter \@secondoftwo
 \fi
}%
\providecommand \natexlab [1]{#1}%
\providecommand \enquote  [1]{``#1''}%
\providecommand \bibnamefont  [1]{#1}%
\providecommand \bibfnamefont [1]{#1}%
\providecommand \citenamefont [1]{#1}%
\providecommand \href@noop [0]{\@secondoftwo}%
\providecommand \href [0]{\begingroup \@sanitize@url \@href}%
\providecommand \@href[1]{\@@startlink{#1}\@@href}%
\providecommand \@@href[1]{\endgroup#1\@@endlink}%
\providecommand \@sanitize@url [0]{\catcode `\\12\catcode `\$12\catcode `\&12\catcode `\#12\catcode `\^12\catcode `\_12\catcode `\%12\relax}%
\providecommand \@@startlink[1]{}%
\providecommand \@@endlink[0]{}%
\providecommand \url  [0]{\begingroup\@sanitize@url \@url }%
\providecommand \@url [1]{\endgroup\@href {#1}{\urlprefix }}%
\providecommand \urlprefix  [0]{URL }%
\providecommand \Eprint [0]{\href }%
\providecommand \doibase [0]{http://dx.doi.org/}%
\providecommand \selectlanguage [0]{\@gobble}%
\providecommand \bibinfo  [0]{\@secondoftwo}%
\providecommand \bibfield  [0]{\@secondoftwo}%
\providecommand \translation [1]{[#1]}%
\providecommand \BibitemOpen [0]{}%
\providecommand \bibitemStop [0]{}%
\providecommand \bibitemNoStop [0]{.\EOS\space}%
\providecommand \EOS [0]{\spacefactor3000\relax}%
\providecommand \BibitemShut  [1]{\csname bibitem#1\endcsname}%
\let\auto@bib@innerbib\@empty
\bibitem [{\citenamefont {\ifmmode~\check{S}\else \v{S}\fi{}mejkal}\ \emph {et~al.}(2022{\natexlab{a}})\citenamefont {\ifmmode~\check{S}\else \v{S}\fi{}mejkal}, \citenamefont {Sinova},\ and\ \citenamefont {Jungwirth}}]{PhysRevX.12.040501}%
  \BibitemOpen
  \bibfield  {author} {\bibinfo {author} {\bibfnamefont {L.}~\bibnamefont {\ifmmode~\check{S}\else \v{S}\fi{}mejkal}}, \bibinfo {author} {\bibfnamefont {J.}~\bibnamefont {Sinova}}, \ and\ \bibinfo {author} {\bibfnamefont {T.}~\bibnamefont {Jungwirth}},\ }\href {\doibase 10.1103/PhysRevX.12.040501} {\bibfield  {journal} {\bibinfo  {journal} {Phys. Rev. X}\ }\textbf {\bibinfo {volume} {12}},\ \bibinfo {pages} {040501} (\bibinfo {year} {2022}{\natexlab{a}})}\BibitemShut {NoStop}%
\bibitem [{\citenamefont {Wu}\ and\ \citenamefont {Zhang}(2004)}]{PhysRevLett.93.036403}%
  \BibitemOpen
  \bibfield  {author} {\bibinfo {author} {\bibfnamefont {C.}~\bibnamefont {Wu}}\ and\ \bibinfo {author} {\bibfnamefont {S.-C.}\ \bibnamefont {Zhang}},\ }\href {\doibase 10.1103/PhysRevLett.93.036403} {\bibfield  {journal} {\bibinfo  {journal} {Phys. Rev. Lett.}\ }\textbf {\bibinfo {volume} {93}},\ \bibinfo {pages} {036403} (\bibinfo {year} {2004})}\BibitemShut {NoStop}%
\bibitem [{\citenamefont {Wu}\ \emph {et~al.}(2007)\citenamefont {Wu}, \citenamefont {Sun}, \citenamefont {Fradkin},\ and\ \citenamefont {Zhang}}]{PhysRevB.75.115103}%
  \BibitemOpen
  \bibfield  {author} {\bibinfo {author} {\bibfnamefont {C.}~\bibnamefont {Wu}}, \bibinfo {author} {\bibfnamefont {K.}~\bibnamefont {Sun}}, \bibinfo {author} {\bibfnamefont {E.}~\bibnamefont {Fradkin}}, \ and\ \bibinfo {author} {\bibfnamefont {S.-C.}\ \bibnamefont {Zhang}},\ }\href {\doibase 10.1103/PhysRevB.75.115103} {\bibfield  {journal} {\bibinfo  {journal} {Phys. Rev. B}\ }\textbf {\bibinfo {volume} {75}},\ \bibinfo {pages} {115103} (\bibinfo {year} {2007})}\BibitemShut {NoStop}%
\bibitem [{\citenamefont {\ifmmode~\check{S}\else \v{S}\fi{}mejkal}\ \emph {et~al.}(2022{\natexlab{b}})\citenamefont {\ifmmode~\check{S}\else \v{S}\fi{}mejkal}, \citenamefont {Sinova},\ and\ \citenamefont {Jungwirth}}]{PhysRevX.12.031042}%
  \BibitemOpen
  \bibfield  {author} {\bibinfo {author} {\bibfnamefont {L.}~\bibnamefont {\ifmmode~\check{S}\else \v{S}\fi{}mejkal}}, \bibinfo {author} {\bibfnamefont {J.}~\bibnamefont {Sinova}}, \ and\ \bibinfo {author} {\bibfnamefont {T.}~\bibnamefont {Jungwirth}},\ }\href {\doibase 10.1103/PhysRevX.12.031042} {\bibfield  {journal} {\bibinfo  {journal} {Phys. Rev. X}\ }\textbf {\bibinfo {volume} {12}},\ \bibinfo {pages} {031042} (\bibinfo {year} {2022}{\natexlab{b}})}\BibitemShut {NoStop}%
\bibitem [{\citenamefont {Chen}\ \emph {et~al.}(2024)\citenamefont {Chen}, \citenamefont {Ren}, \citenamefont {Zhu}, \citenamefont {Yu}, \citenamefont {Zhang}, \citenamefont {Liu}, \citenamefont {Li}, \citenamefont {Liu}, \citenamefont {Li},\ and\ \citenamefont {Liu}}]{PhysRevX.14.031038}%
  \BibitemOpen
  \bibfield  {author} {\bibinfo {author} {\bibfnamefont {X.}~\bibnamefont {Chen}}, \bibinfo {author} {\bibfnamefont {J.}~\bibnamefont {Ren}}, \bibinfo {author} {\bibfnamefont {Y.}~\bibnamefont {Zhu}}, \bibinfo {author} {\bibfnamefont {Y.}~\bibnamefont {Yu}}, \bibinfo {author} {\bibfnamefont {A.}~\bibnamefont {Zhang}}, \bibinfo {author} {\bibfnamefont {P.}~\bibnamefont {Liu}}, \bibinfo {author} {\bibfnamefont {J.}~\bibnamefont {Li}}, \bibinfo {author} {\bibfnamefont {Y.}~\bibnamefont {Liu}}, \bibinfo {author} {\bibfnamefont {C.}~\bibnamefont {Li}}, \ and\ \bibinfo {author} {\bibfnamefont {Q.}~\bibnamefont {Liu}},\ }\href {\doibase 10.1103/PhysRevX.14.031038} {\bibfield  {journal} {\bibinfo  {journal} {Phys. Rev. X}\ }\textbf {\bibinfo {volume} {14}},\ \bibinfo {pages} {031038} (\bibinfo {year} {2024})}\BibitemShut {NoStop}%
\bibitem [{\citenamefont {Li}\ \emph {et~al.}(2022)\citenamefont {Li}, \citenamefont {Yao}, \citenamefont {Wu}, \citenamefont {Hu}, \citenamefont {Gao}, \citenamefont {Wan},\ and\ \citenamefont {Liu}}]{Li:2022aa}%
  \BibitemOpen
  \bibfield  {author} {\bibinfo {author} {\bibfnamefont {J.}~\bibnamefont {Li}}, \bibinfo {author} {\bibfnamefont {Q.}~\bibnamefont {Yao}}, \bibinfo {author} {\bibfnamefont {L.}~\bibnamefont {Wu}}, \bibinfo {author} {\bibfnamefont {Z.}~\bibnamefont {Hu}}, \bibinfo {author} {\bibfnamefont {B.}~\bibnamefont {Gao}}, \bibinfo {author} {\bibfnamefont {X.}~\bibnamefont {Wan}}, \ and\ \bibinfo {author} {\bibfnamefont {Q.}~\bibnamefont {Liu}},\ }\href {\doibase 10.1038/s41467-022-28534-y} {\bibfield  {journal} {\bibinfo  {journal} {Nature Communications}\ }\textbf {\bibinfo {volume} {13}},\ \bibinfo {pages} {919} (\bibinfo {year} {2022})}\BibitemShut {NoStop}%
\bibitem [{\citenamefont {Krempask{\'y}}\ \emph {et~al.}(2024)\citenamefont {Krempask{\'y}}, \citenamefont {{\v S}mejkal}, \citenamefont {D'Souza}, \citenamefont {Hajlaoui}, \citenamefont {Springholz}, \citenamefont {Uhl{\'\i}{\v r}ov{\'a}}, \citenamefont {Alarab}, \citenamefont {Constantinou}, \citenamefont {Strocov}, \citenamefont {Usanov}, \citenamefont {Pudelko}, \citenamefont {Gonz{\'a}lez-Hern{\'a}ndez}, \citenamefont {Birk~Hellenes}, \citenamefont {Jansa}, \citenamefont {Reichlov{\'a}}, \citenamefont {{\v S}ob{\'a}{\v n}}, \citenamefont {Gonzalez~Betancourt}, \citenamefont {Wadley}, \citenamefont {Sinova}, \citenamefont {Kriegner}, \citenamefont {Min{\'a}r}, \citenamefont {Dil},\ and\ \citenamefont {Jungwirth}}]{Krempasky2024-dj}%
  \BibitemOpen
  \bibfield  {author} {\bibinfo {author} {\bibfnamefont {J.}~\bibnamefont {Krempask{\'y}}}, \bibinfo {author} {\bibfnamefont {L.}~\bibnamefont {{\v S}mejkal}}, \bibinfo {author} {\bibfnamefont {S.~W.}\ \bibnamefont {D'Souza}}, \bibinfo {author} {\bibfnamefont {M.}~\bibnamefont {Hajlaoui}}, \bibinfo {author} {\bibfnamefont {G.}~\bibnamefont {Springholz}}, \bibinfo {author} {\bibfnamefont {K.}~\bibnamefont {Uhl{\'\i}{\v r}ov{\'a}}}, \bibinfo {author} {\bibfnamefont {F.}~\bibnamefont {Alarab}}, \bibinfo {author} {\bibfnamefont {P.~C.}\ \bibnamefont {Constantinou}}, \bibinfo {author} {\bibfnamefont {V.}~\bibnamefont {Strocov}}, \bibinfo {author} {\bibfnamefont {D.}~\bibnamefont {Usanov}}, \bibinfo {author} {\bibfnamefont {W.~R.}\ \bibnamefont {Pudelko}}, \bibinfo {author} {\bibfnamefont {R.}~\bibnamefont {Gonz{\'a}lez-Hern{\'a}ndez}}, \bibinfo {author} {\bibfnamefont {A.}~\bibnamefont {Birk~Hellenes}}, \bibinfo {author} {\bibfnamefont {Z.}~\bibnamefont {Jansa}}, \bibinfo {author} {\bibfnamefont {H.}~\bibnamefont
  {Reichlov{\'a}}}, \bibinfo {author} {\bibfnamefont {Z.}~\bibnamefont {{\v S}ob{\'a}{\v n}}}, \bibinfo {author} {\bibfnamefont {R.~D.}\ \bibnamefont {Gonzalez~Betancourt}}, \bibinfo {author} {\bibfnamefont {P.}~\bibnamefont {Wadley}}, \bibinfo {author} {\bibfnamefont {J.}~\bibnamefont {Sinova}}, \bibinfo {author} {\bibfnamefont {D.}~\bibnamefont {Kriegner}}, \bibinfo {author} {\bibfnamefont {J.}~\bibnamefont {Min{\'a}r}}, \bibinfo {author} {\bibfnamefont {J.~H.}\ \bibnamefont {Dil}}, \ and\ \bibinfo {author} {\bibfnamefont {T.}~\bibnamefont {Jungwirth}},\ }\href@noop {} {\bibfield  {journal} {\bibinfo  {journal} {Nature}\ }\textbf {\bibinfo {volume} {626}},\ \bibinfo {pages} {517} (\bibinfo {year} {2024})}\BibitemShut {NoStop}%
\bibitem [{\citenamefont {Chen}\ \emph {et~al.}(2022)\citenamefont {Chen}, \citenamefont {Gu}, \citenamefont {Li}, \citenamefont {Wang},\ and\ \citenamefont {Liu}}]{PhysRevLett.129.276601}%
  \BibitemOpen
  \bibfield  {author} {\bibinfo {author} {\bibfnamefont {W.}~\bibnamefont {Chen}}, \bibinfo {author} {\bibfnamefont {M.}~\bibnamefont {Gu}}, \bibinfo {author} {\bibfnamefont {J.}~\bibnamefont {Li}}, \bibinfo {author} {\bibfnamefont {P.}~\bibnamefont {Wang}}, \ and\ \bibinfo {author} {\bibfnamefont {Q.}~\bibnamefont {Liu}},\ }\href {\doibase 10.1103/PhysRevLett.129.276601} {\bibfield  {journal} {\bibinfo  {journal} {Phys. Rev. Lett.}\ }\textbf {\bibinfo {volume} {129}},\ \bibinfo {pages} {276601} (\bibinfo {year} {2022})}\BibitemShut {NoStop}%
\bibitem [{\citenamefont {Mazin}(2022)}]{mazin2022notes}%
  \BibitemOpen
  \bibfield  {author} {\bibinfo {author} {\bibfnamefont {I.~I.}\ \bibnamefont {Mazin}},\ }\href@noop {} {\enquote {\bibinfo {title} {Notes on altermagnetism and superconductivity},}\ } (\bibinfo {year} {2022}),\ \Eprint {http://arxiv.org/abs/arXiv:2203.05000} {arXiv:2203.05000 [cond-mat.supr-con]} \BibitemShut {NoStop}%
\bibitem [{\citenamefont {Zhang}\ \emph {et~al.}(2024)\citenamefont {Zhang}, \citenamefont {Hu},\ and\ \citenamefont {Neupert}}]{Zhang:2024aa}%
  \BibitemOpen
  \bibfield  {author} {\bibinfo {author} {\bibfnamefont {S.-B.}\ \bibnamefont {Zhang}}, \bibinfo {author} {\bibfnamefont {L.-H.}\ \bibnamefont {Hu}}, \ and\ \bibinfo {author} {\bibfnamefont {T.}~\bibnamefont {Neupert}},\ }\href {\doibase 10.1038/s41467-024-45951-3} {\bibfield  {journal} {\bibinfo  {journal} {Nature Communications}\ }\textbf {\bibinfo {volume} {15}},\ \bibinfo {pages} {1801} (\bibinfo {year} {2024})}\BibitemShut {NoStop}%
\bibitem [{\citenamefont {M\ae{}land}\ \emph {et~al.}(2024)\citenamefont {M\ae{}land}, \citenamefont {Brekke},\ and\ \citenamefont {Sudb\o{}}}]{PhysRevB.109.134515}%
  \BibitemOpen
  \bibfield  {author} {\bibinfo {author} {\bibfnamefont {K.}~\bibnamefont {M\ae{}land}}, \bibinfo {author} {\bibfnamefont {B.}~\bibnamefont {Brekke}}, \ and\ \bibinfo {author} {\bibfnamefont {A.}~\bibnamefont {Sudb\o{}}},\ }\href {\doibase 10.1103/PhysRevB.109.134515} {\bibfield  {journal} {\bibinfo  {journal} {Phys. Rev. B}\ }\textbf {\bibinfo {volume} {109}},\ \bibinfo {pages} {134515} (\bibinfo {year} {2024})}\BibitemShut {NoStop}%
\bibitem [{\citenamefont {Wei}\ \emph {et~al.}(2024)\citenamefont {Wei}, \citenamefont {Xiang}, \citenamefont {Xu}, \citenamefont {Zhang}, \citenamefont {Tang},\ and\ \citenamefont {Wang}}]{PhysRevB.109.L201404}%
  \BibitemOpen
  \bibfield  {author} {\bibinfo {author} {\bibfnamefont {M.}~\bibnamefont {Wei}}, \bibinfo {author} {\bibfnamefont {L.}~\bibnamefont {Xiang}}, \bibinfo {author} {\bibfnamefont {F.}~\bibnamefont {Xu}}, \bibinfo {author} {\bibfnamefont {L.}~\bibnamefont {Zhang}}, \bibinfo {author} {\bibfnamefont {G.}~\bibnamefont {Tang}}, \ and\ \bibinfo {author} {\bibfnamefont {J.}~\bibnamefont {Wang}},\ }\href {\doibase 10.1103/PhysRevB.109.L201404} {\bibfield  {journal} {\bibinfo  {journal} {Phys. Rev. B}\ }\textbf {\bibinfo {volume} {109}},\ \bibinfo {pages} {L201404} (\bibinfo {year} {2024})}\BibitemShut {NoStop}%
\bibitem [{\citenamefont {Chakraborty}\ and\ \citenamefont {Black-Schaffer}(2024)}]{PhysRevB.110.L060508}%
  \BibitemOpen
  \bibfield  {author} {\bibinfo {author} {\bibfnamefont {D.}~\bibnamefont {Chakraborty}}\ and\ \bibinfo {author} {\bibfnamefont {A.~M.}\ \bibnamefont {Black-Schaffer}},\ }\href {\doibase 10.1103/PhysRevB.110.L060508} {\bibfield  {journal} {\bibinfo  {journal} {Phys. Rev. B}\ }\textbf {\bibinfo {volume} {110}},\ \bibinfo {pages} {L060508} (\bibinfo {year} {2024})}\BibitemShut {NoStop}%
\bibitem [{\citenamefont {Sukhachov}\ \emph {et~al.}(2025)\citenamefont {Sukhachov}, \citenamefont {Giil}, \citenamefont {Brekke},\ and\ \citenamefont {Linder}}]{PhysRevB.111.L220403}%
  \BibitemOpen
  \bibfield  {author} {\bibinfo {author} {\bibfnamefont {P.}~\bibnamefont {Sukhachov}}, \bibinfo {author} {\bibfnamefont {H.~G.}\ \bibnamefont {Giil}}, \bibinfo {author} {\bibfnamefont {B.}~\bibnamefont {Brekke}}, \ and\ \bibinfo {author} {\bibfnamefont {J.}~\bibnamefont {Linder}},\ }\href {\doibase 10.1103/PhysRevB.111.L220403} {\bibfield  {journal} {\bibinfo  {journal} {Phys. Rev. B}\ }\textbf {\bibinfo {volume} {111}},\ \bibinfo {pages} {L220403} (\bibinfo {year} {2025})}\BibitemShut {NoStop}%
\bibitem [{\citenamefont {Guo}\ \emph {et~al.}(2010)\citenamefont {Guo}, \citenamefont {Jin}, \citenamefont {Wang}, \citenamefont {Wang}, \citenamefont {Zhu}, \citenamefont {Zhou}, \citenamefont {He},\ and\ \citenamefont {Chen}}]{PhysRevB.82.180520}%
  \BibitemOpen
  \bibfield  {author} {\bibinfo {author} {\bibfnamefont {J.}~\bibnamefont {Guo}}, \bibinfo {author} {\bibfnamefont {S.}~\bibnamefont {Jin}}, \bibinfo {author} {\bibfnamefont {G.}~\bibnamefont {Wang}}, \bibinfo {author} {\bibfnamefont {S.}~\bibnamefont {Wang}}, \bibinfo {author} {\bibfnamefont {K.}~\bibnamefont {Zhu}}, \bibinfo {author} {\bibfnamefont {T.}~\bibnamefont {Zhou}}, \bibinfo {author} {\bibfnamefont {M.}~\bibnamefont {He}}, \ and\ \bibinfo {author} {\bibfnamefont {X.}~\bibnamefont {Chen}},\ }\href {\doibase 10.1103/PhysRevB.82.180520} {\bibfield  {journal} {\bibinfo  {journal} {Phys. Rev. B}\ }\textbf {\bibinfo {volume} {82}},\ \bibinfo {pages} {180520} (\bibinfo {year} {2010})}\BibitemShut {NoStop}%
\bibitem [{\citenamefont {Wang}\ \emph {et~al.}(2011)\citenamefont {Wang}, \citenamefont {Dong}, \citenamefont {Li}, \citenamefont {Mao}, \citenamefont {Zhu}, \citenamefont {Feng}, \citenamefont {Yuan},\ and\ \citenamefont {Fang}}]{Wang_2011}%
  \BibitemOpen
  \bibfield  {author} {\bibinfo {author} {\bibfnamefont {H.-D.}\ \bibnamefont {Wang}}, \bibinfo {author} {\bibfnamefont {C.-H.}\ \bibnamefont {Dong}}, \bibinfo {author} {\bibfnamefont {Z.-J.}\ \bibnamefont {Li}}, \bibinfo {author} {\bibfnamefont {Q.-H.}\ \bibnamefont {Mao}}, \bibinfo {author} {\bibfnamefont {S.-S.}\ \bibnamefont {Zhu}}, \bibinfo {author} {\bibfnamefont {C.-M.}\ \bibnamefont {Feng}}, \bibinfo {author} {\bibfnamefont {H.~Q.}\ \bibnamefont {Yuan}}, \ and\ \bibinfo {author} {\bibfnamefont {M.-H.}\ \bibnamefont {Fang}},\ }\href {\doibase 10.1209/0295-5075/93/47004} {\bibfield  {journal} {\bibinfo  {journal} {Europhysics Letters}\ }\textbf {\bibinfo {volume} {93}},\ \bibinfo {pages} {47004} (\bibinfo {year} {2011})}\BibitemShut {NoStop}%
\bibitem [{\citenamefont {Fang}\ \emph {et~al.}(2011)\citenamefont {Fang}, \citenamefont {Wang}, \citenamefont {Dong}, \citenamefont {Li}, \citenamefont {Feng}, \citenamefont {Chen},\ and\ \citenamefont {Yuan}}]{Fang_2011}%
  \BibitemOpen
  \bibfield  {author} {\bibinfo {author} {\bibfnamefont {M.-H.}\ \bibnamefont {Fang}}, \bibinfo {author} {\bibfnamefont {H.-D.}\ \bibnamefont {Wang}}, \bibinfo {author} {\bibfnamefont {C.-H.}\ \bibnamefont {Dong}}, \bibinfo {author} {\bibfnamefont {Z.-J.}\ \bibnamefont {Li}}, \bibinfo {author} {\bibfnamefont {C.-M.}\ \bibnamefont {Feng}}, \bibinfo {author} {\bibfnamefont {J.}~\bibnamefont {Chen}}, \ and\ \bibinfo {author} {\bibfnamefont {H.~Q.}\ \bibnamefont {Yuan}},\ }\href {\doibase 10.1209/0295-5075/94/27009} {\bibfield  {journal} {\bibinfo  {journal} {Europhysics Letters}\ }\textbf {\bibinfo {volume} {94}},\ \bibinfo {pages} {27009} (\bibinfo {year} {2011})}\BibitemShut {NoStop}%
\bibitem [{\citenamefont {Sasmal}\ \emph {et~al.}(2008)\citenamefont {Sasmal}, \citenamefont {Lv}, \citenamefont {Lorenz}, \citenamefont {Guloy}, \citenamefont {Chen}, \citenamefont {Xue},\ and\ \citenamefont {Chu}}]{PhysRevLett.101.107007}%
  \BibitemOpen
  \bibfield  {author} {\bibinfo {author} {\bibfnamefont {K.}~\bibnamefont {Sasmal}}, \bibinfo {author} {\bibfnamefont {B.}~\bibnamefont {Lv}}, \bibinfo {author} {\bibfnamefont {B.}~\bibnamefont {Lorenz}}, \bibinfo {author} {\bibfnamefont {A.~M.}\ \bibnamefont {Guloy}}, \bibinfo {author} {\bibfnamefont {F.}~\bibnamefont {Chen}}, \bibinfo {author} {\bibfnamefont {Y.-Y.}\ \bibnamefont {Xue}}, \ and\ \bibinfo {author} {\bibfnamefont {C.-W.}\ \bibnamefont {Chu}},\ }\href {\doibase 10.1103/PhysRevLett.101.107007} {\bibfield  {journal} {\bibinfo  {journal} {Phys. Rev. Lett.}\ }\textbf {\bibinfo {volume} {101}},\ \bibinfo {pages} {107007} (\bibinfo {year} {2008})}\BibitemShut {NoStop}%
\bibitem [{\citenamefont {Bao}\ \emph {et~al.}(2011)\citenamefont {Bao}, \citenamefont {Huang}, \citenamefont {Chen}, \citenamefont {Wang}, \citenamefont {He},\ and\ \citenamefont {Qiu}}]{Bao_2011}%
  \BibitemOpen
  \bibfield  {author} {\bibinfo {author} {\bibfnamefont {W.}~\bibnamefont {Bao}}, \bibinfo {author} {\bibfnamefont {Q.-Z.}\ \bibnamefont {Huang}}, \bibinfo {author} {\bibfnamefont {G.-F.}\ \bibnamefont {Chen}}, \bibinfo {author} {\bibfnamefont {D.-M.}\ \bibnamefont {Wang}}, \bibinfo {author} {\bibfnamefont {J.-B.}\ \bibnamefont {He}}, \ and\ \bibinfo {author} {\bibfnamefont {Y.-M.}\ \bibnamefont {Qiu}},\ }\href {\doibase 10.1088/0256-307X/28/8/086104} {\bibfield  {journal} {\bibinfo  {journal} {Chinese Physics Letters}\ }\textbf {\bibinfo {volume} {28}},\ \bibinfo {pages} {086104} (\bibinfo {year} {2011})}\BibitemShut {NoStop}%
\bibitem [{\citenamefont {Cao}\ and\ \citenamefont {Dai}(2011)}]{PhysRevLett.107.056401}%
  \BibitemOpen
  \bibfield  {author} {\bibinfo {author} {\bibfnamefont {C.}~\bibnamefont {Cao}}\ and\ \bibinfo {author} {\bibfnamefont {J.}~\bibnamefont {Dai}},\ }\href {\doibase 10.1103/PhysRevLett.107.056401} {\bibfield  {journal} {\bibinfo  {journal} {Phys. Rev. Lett.}\ }\textbf {\bibinfo {volume} {107}},\ \bibinfo {pages} {056401} (\bibinfo {year} {2011})}\BibitemShut {NoStop}%
\bibitem [{\citenamefont {Yan}\ \emph {et~al.}(2011)\citenamefont {Yan}, \citenamefont {Gao}, \citenamefont {Lu},\ and\ \citenamefont {Xiang}}]{PhysRevB.83.233205}%
  \BibitemOpen
  \bibfield  {author} {\bibinfo {author} {\bibfnamefont {X.-W.}\ \bibnamefont {Yan}}, \bibinfo {author} {\bibfnamefont {M.}~\bibnamefont {Gao}}, \bibinfo {author} {\bibfnamefont {Z.-Y.}\ \bibnamefont {Lu}}, \ and\ \bibinfo {author} {\bibfnamefont {T.}~\bibnamefont {Xiang}},\ }\href {\doibase 10.1103/PhysRevB.83.233205} {\bibfield  {journal} {\bibinfo  {journal} {Phys. Rev. B}\ }\textbf {\bibinfo {volume} {83}},\ \bibinfo {pages} {233205} (\bibinfo {year} {2011})}\BibitemShut {NoStop}%
\bibitem [{\citenamefont {Ye}\ \emph {et~al.}(2011)\citenamefont {Ye}, \citenamefont {Chi}, \citenamefont {Bao}, \citenamefont {Wang}, \citenamefont {Ying}, \citenamefont {Chen}, \citenamefont {Wang}, \citenamefont {Dong},\ and\ \citenamefont {Fang}}]{PhysRevLett.107.137003}%
  \BibitemOpen
  \bibfield  {author} {\bibinfo {author} {\bibfnamefont {F.}~\bibnamefont {Ye}}, \bibinfo {author} {\bibfnamefont {S.}~\bibnamefont {Chi}}, \bibinfo {author} {\bibfnamefont {W.}~\bibnamefont {Bao}}, \bibinfo {author} {\bibfnamefont {X.~F.}\ \bibnamefont {Wang}}, \bibinfo {author} {\bibfnamefont {J.~J.}\ \bibnamefont {Ying}}, \bibinfo {author} {\bibfnamefont {X.~H.}\ \bibnamefont {Chen}}, \bibinfo {author} {\bibfnamefont {H.~D.}\ \bibnamefont {Wang}}, \bibinfo {author} {\bibfnamefont {C.~H.}\ \bibnamefont {Dong}}, \ and\ \bibinfo {author} {\bibfnamefont {M.}~\bibnamefont {Fang}},\ }\href {\doibase 10.1103/PhysRevLett.107.137003} {\bibfield  {journal} {\bibinfo  {journal} {Phys. Rev. Lett.}\ }\textbf {\bibinfo {volume} {107}},\ \bibinfo {pages} {137003} (\bibinfo {year} {2011})}\BibitemShut {NoStop}%
\bibitem [{\citenamefont {Toulemonde}\ \emph {et~al.}(2013)\citenamefont {Toulemonde}, \citenamefont {Santos-Cottin}, \citenamefont {Lepoittevin}, \citenamefont {Strobel},\ and\ \citenamefont {Marcus}}]{Toulemonde_2013}%
  \BibitemOpen
  \bibfield  {author} {\bibinfo {author} {\bibfnamefont {P.}~\bibnamefont {Toulemonde}}, \bibinfo {author} {\bibfnamefont {D.}~\bibnamefont {Santos-Cottin}}, \bibinfo {author} {\bibfnamefont {C.}~\bibnamefont {Lepoittevin}}, \bibinfo {author} {\bibfnamefont {P.}~\bibnamefont {Strobel}}, \ and\ \bibinfo {author} {\bibfnamefont {J.}~\bibnamefont {Marcus}},\ }\href {\doibase 10.1088/0953-8984/25/7/075703} {\bibfield  {journal} {\bibinfo  {journal} {Journal of Physics: Condensed Matter}\ }\textbf {\bibinfo {volume} {25}},\ \bibinfo {pages} {075703} (\bibinfo {year} {2013})}\BibitemShut {NoStop}%
\bibitem [{\citenamefont {Bao}(2014)}]{Bao_2015}%
  \BibitemOpen
  \bibfield  {author} {\bibinfo {author} {\bibfnamefont {W.}~\bibnamefont {Bao}},\ }\href {\doibase 10.1088/0953-8984/27/2/023201} {\bibfield  {journal} {\bibinfo  {journal} {Journal of Physics: Condensed Matter}\ }\textbf {\bibinfo {volume} {27}},\ \bibinfo {pages} {023201} (\bibinfo {year} {2014})}\BibitemShut {NoStop}%
\bibitem [{\citenamefont {Mangelis}\ \emph {et~al.}(2019)\citenamefont {Mangelis}, \citenamefont {Koch}, \citenamefont {Lei}, \citenamefont {Neder}, \citenamefont {McDonnell}, \citenamefont {Feygenson}, \citenamefont {Petrovic}, \citenamefont {Lappas},\ and\ \citenamefont {Bozin}}]{PhysRevB.100.094108}%
  \BibitemOpen
  \bibfield  {author} {\bibinfo {author} {\bibfnamefont {P.}~\bibnamefont {Mangelis}}, \bibinfo {author} {\bibfnamefont {R.~J.}\ \bibnamefont {Koch}}, \bibinfo {author} {\bibfnamefont {H.}~\bibnamefont {Lei}}, \bibinfo {author} {\bibfnamefont {R.~B.}\ \bibnamefont {Neder}}, \bibinfo {author} {\bibfnamefont {M.~T.}\ \bibnamefont {McDonnell}}, \bibinfo {author} {\bibfnamefont {M.}~\bibnamefont {Feygenson}}, \bibinfo {author} {\bibfnamefont {C.}~\bibnamefont {Petrovic}}, \bibinfo {author} {\bibfnamefont {A.}~\bibnamefont {Lappas}}, \ and\ \bibinfo {author} {\bibfnamefont {E.~S.}\ \bibnamefont {Bozin}},\ }\href {\doibase 10.1103/PhysRevB.100.094108} {\bibfield  {journal} {\bibinfo  {journal} {Phys. Rev. B}\ }\textbf {\bibinfo {volume} {100}},\ \bibinfo {pages} {094108} (\bibinfo {year} {2019})}\BibitemShut {NoStop}%
\bibitem [{\citenamefont {Croitori}\ \emph {et~al.}(2020)\citenamefont {Croitori}, \citenamefont {Filippova}, \citenamefont {Kravtsov}, \citenamefont {G\"unther}, \citenamefont {Widmann}, \citenamefont {Reuter}, \citenamefont {Krug~von Nidda}, \citenamefont {Deisenhofer}, \citenamefont {Loidl},\ and\ \citenamefont {Tsurkan}}]{PhysRevB.101.054516}%
  \BibitemOpen
  \bibfield  {author} {\bibinfo {author} {\bibfnamefont {D.}~\bibnamefont {Croitori}}, \bibinfo {author} {\bibfnamefont {I.}~\bibnamefont {Filippova}}, \bibinfo {author} {\bibfnamefont {V.}~\bibnamefont {Kravtsov}}, \bibinfo {author} {\bibfnamefont {A.}~\bibnamefont {G\"unther}}, \bibinfo {author} {\bibfnamefont {S.}~\bibnamefont {Widmann}}, \bibinfo {author} {\bibfnamefont {D.}~\bibnamefont {Reuter}}, \bibinfo {author} {\bibfnamefont {H.-A.}\ \bibnamefont {Krug~von Nidda}}, \bibinfo {author} {\bibfnamefont {J.}~\bibnamefont {Deisenhofer}}, \bibinfo {author} {\bibfnamefont {A.}~\bibnamefont {Loidl}}, \ and\ \bibinfo {author} {\bibfnamefont {V.}~\bibnamefont {Tsurkan}},\ }\href {\doibase 10.1103/PhysRevB.101.054516} {\bibfield  {journal} {\bibinfo  {journal} {Phys. Rev. B}\ }\textbf {\bibinfo {volume} {101}},\ \bibinfo {pages} {054516} (\bibinfo {year} {2020})}\BibitemShut {NoStop}%
\bibitem [{\citenamefont {Chen}\ \emph {et~al.}(2021)\citenamefont {Chen}, \citenamefont {Jiang}, \citenamefont {Yang}, \citenamefont {Dudin}, \citenamefont {Barinov}, \citenamefont {Liu}, \citenamefont {Wen}, \citenamefont {Yang},\ and\ \citenamefont {Chen}}]{Chen:2021aa}%
  \BibitemOpen
  \bibfield  {author} {\bibinfo {author} {\bibfnamefont {Y.}~\bibnamefont {Chen}}, \bibinfo {author} {\bibfnamefont {J.}~\bibnamefont {Jiang}}, \bibinfo {author} {\bibfnamefont {H.}~\bibnamefont {Yang}}, \bibinfo {author} {\bibfnamefont {P.}~\bibnamefont {Dudin}}, \bibinfo {author} {\bibfnamefont {A.}~\bibnamefont {Barinov}}, \bibinfo {author} {\bibfnamefont {Z.}~\bibnamefont {Liu}}, \bibinfo {author} {\bibfnamefont {H.}~\bibnamefont {Wen}}, \bibinfo {author} {\bibfnamefont {L.}~\bibnamefont {Yang}}, \ and\ \bibinfo {author} {\bibfnamefont {Y.}~\bibnamefont {Chen}},\ }\href {\doibase 10.1007/s12274-020-3119-8} {\bibfield  {journal} {\bibinfo  {journal} {Nano Research}\ }\textbf {\bibinfo {volume} {14}},\ \bibinfo {pages} {823} (\bibinfo {year} {2021})}\BibitemShut {NoStop}%
\bibitem [{\citenamefont {Chen}\ \emph {et~al.}(2014)\citenamefont {Chen}, \citenamefont {Chang}, \citenamefont {Chang}, \citenamefont {Fang}, \citenamefont {Wang}, \citenamefont {Chao}, \citenamefont {Tseng}, \citenamefont {Lee}, \citenamefont {Wu}, \citenamefont {Wen}, \citenamefont {Tang}, \citenamefont {Chen}, \citenamefont {Wang}, \citenamefont {Wu},\ and\ \citenamefont {Dyck}}]{pnas.1321160111}%
  \BibitemOpen
  \bibfield  {author} {\bibinfo {author} {\bibfnamefont {T.-K.}\ \bibnamefont {Chen}}, \bibinfo {author} {\bibfnamefont {C.-C.}\ \bibnamefont {Chang}}, \bibinfo {author} {\bibfnamefont {H.-H.}\ \bibnamefont {Chang}}, \bibinfo {author} {\bibfnamefont {A.-H.}\ \bibnamefont {Fang}}, \bibinfo {author} {\bibfnamefont {C.-H.}\ \bibnamefont {Wang}}, \bibinfo {author} {\bibfnamefont {W.-H.}\ \bibnamefont {Chao}}, \bibinfo {author} {\bibfnamefont {C.-M.}\ \bibnamefont {Tseng}}, \bibinfo {author} {\bibfnamefont {Y.-C.}\ \bibnamefont {Lee}}, \bibinfo {author} {\bibfnamefont {Y.-R.}\ \bibnamefont {Wu}}, \bibinfo {author} {\bibfnamefont {M.-H.}\ \bibnamefont {Wen}}, \bibinfo {author} {\bibfnamefont {H.-Y.}\ \bibnamefont {Tang}}, \bibinfo {author} {\bibfnamefont {F.-R.}\ \bibnamefont {Chen}}, \bibinfo {author} {\bibfnamefont {M.-J.}\ \bibnamefont {Wang}}, \bibinfo {author} {\bibfnamefont {M.-K.}\ \bibnamefont {Wu}}, \ and\ \bibinfo {author} {\bibfnamefont {D.~V.}\ \bibnamefont {Dyck}},\ }\href {\doibase
  10.1073/pnas.1321160111} {\bibfield  {journal} {\bibinfo  {journal} {Proceedings of the National Academy of Sciences}\ }\textbf {\bibinfo {volume} {111}},\ \bibinfo {pages} {63} (\bibinfo {year} {2014})},\ \Eprint {http://arxiv.org/abs/https://www.pnas.org/doi/pdf/10.1073/pnas.1321160111} {https://www.pnas.org/doi/pdf/10.1073/pnas.1321160111} \BibitemShut {NoStop}%
\bibitem [{\citenamefont {Gao}\ \emph {et~al.}(2017)\citenamefont {Gao}, \citenamefont {Kong}, \citenamefont {Yan}, \citenamefont {Lu},\ and\ \citenamefont {Xiang}}]{PhysRevB.95.174523}%
  \BibitemOpen
  \bibfield  {author} {\bibinfo {author} {\bibfnamefont {M.}~\bibnamefont {Gao}}, \bibinfo {author} {\bibfnamefont {X.}~\bibnamefont {Kong}}, \bibinfo {author} {\bibfnamefont {X.-W.}\ \bibnamefont {Yan}}, \bibinfo {author} {\bibfnamefont {Z.-Y.}\ \bibnamefont {Lu}}, \ and\ \bibinfo {author} {\bibfnamefont {T.}~\bibnamefont {Xiang}},\ }\href {\doibase 10.1103/PhysRevB.95.174523} {\bibfield  {journal} {\bibinfo  {journal} {Phys. Rev. B}\ }\textbf {\bibinfo {volume} {95}},\ \bibinfo {pages} {174523} (\bibinfo {year} {2017})}\BibitemShut {NoStop}%
\bibitem [{\citenamefont {Yeh}\ \emph {et~al.}(2020)\citenamefont {Yeh}, \citenamefont {Chen}, \citenamefont {Lo}, \citenamefont {Wu}, \citenamefont {Wang}, \citenamefont {Chang-Liao},\ and\ \citenamefont {Wu}}]{fphy.2020.567054}%
  \BibitemOpen
  \bibfield  {author} {\bibinfo {author} {\bibfnamefont {K.-Y.}\ \bibnamefont {Yeh}}, \bibinfo {author} {\bibfnamefont {Y.-R.}\ \bibnamefont {Chen}}, \bibinfo {author} {\bibfnamefont {T.-S.}\ \bibnamefont {Lo}}, \bibinfo {author} {\bibfnamefont {P.~M.}\ \bibnamefont {Wu}}, \bibinfo {author} {\bibfnamefont {M.-J.}\ \bibnamefont {Wang}}, \bibinfo {author} {\bibfnamefont {K.-S.}\ \bibnamefont {Chang-Liao}}, \ and\ \bibinfo {author} {\bibfnamefont {M.-K.}\ \bibnamefont {Wu}},\ }\href {\doibase 10.3389/fphy.2020.567054} {\bibfield  {journal} {\bibinfo  {journal} {Frontiers in Physics}\ }\textbf {\bibinfo {volume} {8}} (\bibinfo {year} {2020}),\ 10.3389/fphy.2020.567054}\BibitemShut {NoStop}%
\bibitem [{\citenamefont {Sun}\ \emph {et~al.}(2012)\citenamefont {Sun}, \citenamefont {Chen}, \citenamefont {Guo}, \citenamefont {Gao}, \citenamefont {Huang}, \citenamefont {Wang}, \citenamefont {Fang}, \citenamefont {Chen}, \citenamefont {Chen}, \citenamefont {Wu}, \citenamefont {Zhang}, \citenamefont {Gu}, \citenamefont {Dong}, \citenamefont {Wang}, \citenamefont {Yang}, \citenamefont {Li}, \citenamefont {Dai}, \citenamefont {Mao},\ and\ \citenamefont {Zhao}}]{Sun:2012aa}%
  \BibitemOpen
  \bibfield  {author} {\bibinfo {author} {\bibfnamefont {L.}~\bibnamefont {Sun}}, \bibinfo {author} {\bibfnamefont {X.-J.}\ \bibnamefont {Chen}}, \bibinfo {author} {\bibfnamefont {J.}~\bibnamefont {Guo}}, \bibinfo {author} {\bibfnamefont {P.}~\bibnamefont {Gao}}, \bibinfo {author} {\bibfnamefont {Q.-Z.}\ \bibnamefont {Huang}}, \bibinfo {author} {\bibfnamefont {H.}~\bibnamefont {Wang}}, \bibinfo {author} {\bibfnamefont {M.}~\bibnamefont {Fang}}, \bibinfo {author} {\bibfnamefont {X.}~\bibnamefont {Chen}}, \bibinfo {author} {\bibfnamefont {G.}~\bibnamefont {Chen}}, \bibinfo {author} {\bibfnamefont {Q.}~\bibnamefont {Wu}}, \bibinfo {author} {\bibfnamefont {C.}~\bibnamefont {Zhang}}, \bibinfo {author} {\bibfnamefont {D.}~\bibnamefont {Gu}}, \bibinfo {author} {\bibfnamefont {X.}~\bibnamefont {Dong}}, \bibinfo {author} {\bibfnamefont {L.}~\bibnamefont {Wang}}, \bibinfo {author} {\bibfnamefont {K.}~\bibnamefont {Yang}}, \bibinfo {author} {\bibfnamefont {A.}~\bibnamefont {Li}}, \bibinfo {author} {\bibfnamefont
  {X.}~\bibnamefont {Dai}}, \bibinfo {author} {\bibfnamefont {H.-k.}\ \bibnamefont {Mao}}, \ and\ \bibinfo {author} {\bibfnamefont {Z.}~\bibnamefont {Zhao}},\ }\href {\doibase 10.1038/nature10813} {\bibfield  {journal} {\bibinfo  {journal} {Nature}\ }\textbf {\bibinfo {volume} {483}},\ \bibinfo {pages} {67} (\bibinfo {year} {2012})}\BibitemShut {NoStop}%
\bibitem [{\citenamefont {Ksenofontov}\ \emph {et~al.}(2012)\citenamefont {Ksenofontov}, \citenamefont {Medvedev}, \citenamefont {Schoop}, \citenamefont {Wortmann}, \citenamefont {Palasyuk}, \citenamefont {Tsurkan}, \citenamefont {Deisenhofer}, \citenamefont {Loidl},\ and\ \citenamefont {Felser}}]{PhysRevB.85.214519}%
  \BibitemOpen
  \bibfield  {author} {\bibinfo {author} {\bibfnamefont {V.}~\bibnamefont {Ksenofontov}}, \bibinfo {author} {\bibfnamefont {S.~A.}\ \bibnamefont {Medvedev}}, \bibinfo {author} {\bibfnamefont {L.~M.}\ \bibnamefont {Schoop}}, \bibinfo {author} {\bibfnamefont {G.}~\bibnamefont {Wortmann}}, \bibinfo {author} {\bibfnamefont {T.}~\bibnamefont {Palasyuk}}, \bibinfo {author} {\bibfnamefont {V.}~\bibnamefont {Tsurkan}}, \bibinfo {author} {\bibfnamefont {J.}~\bibnamefont {Deisenhofer}}, \bibinfo {author} {\bibfnamefont {A.}~\bibnamefont {Loidl}}, \ and\ \bibinfo {author} {\bibfnamefont {C.}~\bibnamefont {Felser}},\ }\href {\doibase 10.1103/PhysRevB.85.214519} {\bibfield  {journal} {\bibinfo  {journal} {Phys. Rev. B}\ }\textbf {\bibinfo {volume} {85}},\ \bibinfo {pages} {214519} (\bibinfo {year} {2012})}\BibitemShut {NoStop}%
\bibitem [{\citenamefont {Chi}\ \emph {et~al.}(2013)\citenamefont {Chi}, \citenamefont {Ye}, \citenamefont {Bao}, \citenamefont {Fang}, \citenamefont {Wang}, \citenamefont {Dong}, \citenamefont {Savici}, \citenamefont {Granroth}, \citenamefont {Stone},\ and\ \citenamefont {Fishman}}]{PhysRevB.87.100501}%
  \BibitemOpen
  \bibfield  {author} {\bibinfo {author} {\bibfnamefont {S.}~\bibnamefont {Chi}}, \bibinfo {author} {\bibfnamefont {F.}~\bibnamefont {Ye}}, \bibinfo {author} {\bibfnamefont {W.}~\bibnamefont {Bao}}, \bibinfo {author} {\bibfnamefont {M.}~\bibnamefont {Fang}}, \bibinfo {author} {\bibfnamefont {H.~D.}\ \bibnamefont {Wang}}, \bibinfo {author} {\bibfnamefont {C.~H.}\ \bibnamefont {Dong}}, \bibinfo {author} {\bibfnamefont {A.~T.}\ \bibnamefont {Savici}}, \bibinfo {author} {\bibfnamefont {G.~E.}\ \bibnamefont {Granroth}}, \bibinfo {author} {\bibfnamefont {M.~B.}\ \bibnamefont {Stone}}, \ and\ \bibinfo {author} {\bibfnamefont {R.~S.}\ \bibnamefont {Fishman}},\ }\href {\doibase 10.1103/PhysRevB.87.100501} {\bibfield  {journal} {\bibinfo  {journal} {Phys. Rev. B}\ }\textbf {\bibinfo {volume} {87}},\ \bibinfo {pages} {100501} (\bibinfo {year} {2013})}\BibitemShut {NoStop}%
\bibitem [{\citenamefont {Cao}\ \emph {et~al.}(2011)\citenamefont {Cao}, \citenamefont {Fang},\ and\ \citenamefont {Dai}}]{cao2011blockspinmagneticphase}%
  \BibitemOpen
  \bibfield  {author} {\bibinfo {author} {\bibfnamefont {C.}~\bibnamefont {Cao}}, \bibinfo {author} {\bibfnamefont {M.}~\bibnamefont {Fang}}, \ and\ \bibinfo {author} {\bibfnamefont {J.}~\bibnamefont {Dai}},\ }\href {https://arxiv.org/abs/1108.4322} {\enquote {\bibinfo {title} {Block spin magnetic phase transition of a$_y$fe$_{1.6}$se$_2$ under high pressure},}\ } (\bibinfo {year} {2011}),\ \Eprint {http://arxiv.org/abs/1108.4322} {arXiv:1108.4322 [cond-mat.supr-con]} \BibitemShut {NoStop}%
\bibitem [{\citenamefont {Kotliar}\ \emph {et~al.}(2006)\citenamefont {Kotliar}, \citenamefont {Savrasov}, \citenamefont {Haule}, \citenamefont {Oudovenko}, \citenamefont {Parcollet},\ and\ \citenamefont {Marianetti}}]{method:dmft2}%
  \BibitemOpen
  \bibfield  {author} {\bibinfo {author} {\bibfnamefont {G.}~\bibnamefont {Kotliar}}, \bibinfo {author} {\bibfnamefont {S.~Y.}\ \bibnamefont {Savrasov}}, \bibinfo {author} {\bibfnamefont {K.}~\bibnamefont {Haule}}, \bibinfo {author} {\bibfnamefont {V.~S.}\ \bibnamefont {Oudovenko}}, \bibinfo {author} {\bibfnamefont {O.}~\bibnamefont {Parcollet}}, \ and\ \bibinfo {author} {\bibfnamefont {C.~A.}\ \bibnamefont {Marianetti}},\ }\href {\doibase 10.1103/RevModPhys.78.865} {\bibfield  {journal} {\bibinfo  {journal} {Rev. Mod. Phys.}\ }\textbf {\bibinfo {volume} {78}},\ \bibinfo {pages} {865} (\bibinfo {year} {2006})}\BibitemShut {NoStop}%
\bibitem [{\citenamefont {Haule}(2007)}]{method:ctqmc_dmft}%
  \BibitemOpen
  \bibfield  {author} {\bibinfo {author} {\bibfnamefont {K.}~\bibnamefont {Haule}},\ }\href {\doibase 10.1103/PhysRevB.75.155113} {\bibfield  {journal} {\bibinfo  {journal} {Phys. Rev. B}\ }\textbf {\bibinfo {volume} {75}},\ \bibinfo {pages} {155113} (\bibinfo {year} {2007})}\BibitemShut {NoStop}%
\bibitem [{\citenamefont {{\v S}mejkal}\ \emph {et~al.}(2020)\citenamefont {{\v S}mejkal}, \citenamefont {Gonz{\'a}lez-Hern{\'a}ndez}, \citenamefont {Jungwirth},\ and\ \citenamefont {Sinova}}]{Smejkal:aa}%
  \BibitemOpen
  \bibfield  {author} {\bibinfo {author} {\bibfnamefont {L.}~\bibnamefont {{\v S}mejkal}}, \bibinfo {author} {\bibfnamefont {R.}~\bibnamefont {Gonz{\'a}lez-Hern{\'a}ndez}}, \bibinfo {author} {\bibfnamefont {T.}~\bibnamefont {Jungwirth}}, \ and\ \bibinfo {author} {\bibfnamefont {J.}~\bibnamefont {Sinova}},\ }\href {\doibase 10.1126/sciadv.aaz8809} {\bibfield  {journal} {\bibinfo  {journal} {Science Advances}\ }\textbf {\bibinfo {volume} {6}},\ \bibinfo {pages} {eaaz8809} (\bibinfo {year} {2020})}\BibitemShut {NoStop}%
\bibitem [{\citenamefont {Shao}\ \emph {et~al.}(2021)\citenamefont {Shao}, \citenamefont {Zhang}, \citenamefont {Li}, \citenamefont {Eom},\ and\ \citenamefont {Tsymbal}}]{Shao:2021aa}%
  \BibitemOpen
  \bibfield  {author} {\bibinfo {author} {\bibfnamefont {D.-F.}\ \bibnamefont {Shao}}, \bibinfo {author} {\bibfnamefont {S.-H.}\ \bibnamefont {Zhang}}, \bibinfo {author} {\bibfnamefont {M.}~\bibnamefont {Li}}, \bibinfo {author} {\bibfnamefont {C.-B.}\ \bibnamefont {Eom}}, \ and\ \bibinfo {author} {\bibfnamefont {E.~Y.}\ \bibnamefont {Tsymbal}},\ }\href {\doibase 10.1038/s41467-021-26915-3} {\bibfield  {journal} {\bibinfo  {journal} {Nature Communications}\ }\textbf {\bibinfo {volume} {12}},\ \bibinfo {pages} {7061} (\bibinfo {year} {2021})}\BibitemShut {NoStop}%
\bibitem [{\citenamefont {Xu}\ \emph {et~al.}(2025)\citenamefont {Xu}, \citenamefont {Wu}, \citenamefont {Zhi}, \citenamefont {Cao}, \citenamefont {Dai}, \citenamefont {Cao}, \citenamefont {Wang},\ and\ \citenamefont {Lin}}]{Xu:2025aa}%
  \BibitemOpen
  \bibfield  {author} {\bibinfo {author} {\bibfnamefont {C.}~\bibnamefont {Xu}}, \bibinfo {author} {\bibfnamefont {S.}~\bibnamefont {Wu}}, \bibinfo {author} {\bibfnamefont {G.-X.}\ \bibnamefont {Zhi}}, \bibinfo {author} {\bibfnamefont {G.}~\bibnamefont {Cao}}, \bibinfo {author} {\bibfnamefont {J.}~\bibnamefont {Dai}}, \bibinfo {author} {\bibfnamefont {C.}~\bibnamefont {Cao}}, \bibinfo {author} {\bibfnamefont {X.}~\bibnamefont {Wang}}, \ and\ \bibinfo {author} {\bibfnamefont {H.-Q.}\ \bibnamefont {Lin}},\ }\href {\doibase 10.1038/s41467-025-58446-6} {\bibfield  {journal} {\bibinfo  {journal} {Nature Communications}\ }\textbf {\bibinfo {volume} {16}},\ \bibinfo {pages} {3114} (\bibinfo {year} {2025})}\BibitemShut {NoStop}%
\bibitem [{\citenamefont {Ke}\ \emph {et~al.}(2012)\citenamefont {Ke}, \citenamefont {van Schilfgaarde},\ and\ \citenamefont {Antropov}}]{PhysRevB.86.020402}%
  \BibitemOpen
  \bibfield  {author} {\bibinfo {author} {\bibfnamefont {L.}~\bibnamefont {Ke}}, \bibinfo {author} {\bibfnamefont {M.}~\bibnamefont {van Schilfgaarde}}, \ and\ \bibinfo {author} {\bibfnamefont {V.}~\bibnamefont {Antropov}},\ }\href {\doibase 10.1103/PhysRevB.86.020402} {\bibfield  {journal} {\bibinfo  {journal} {Phys. Rev. B}\ }\textbf {\bibinfo {volume} {86}},\ \bibinfo {pages} {020402} (\bibinfo {year} {2012})}\BibitemShut {NoStop}%
\bibitem [{\citenamefont {Mazin}\ \emph {et~al.}(2008)\citenamefont {Mazin}, \citenamefont {Johannes}, \citenamefont {Boeri}, \citenamefont {Koepernik},\ and\ \citenamefont {Singh}}]{PhysRevB.78.085104}%
  \BibitemOpen
  \bibfield  {author} {\bibinfo {author} {\bibfnamefont {I.~I.}\ \bibnamefont {Mazin}}, \bibinfo {author} {\bibfnamefont {M.~D.}\ \bibnamefont {Johannes}}, \bibinfo {author} {\bibfnamefont {L.}~\bibnamefont {Boeri}}, \bibinfo {author} {\bibfnamefont {K.}~\bibnamefont {Koepernik}}, \ and\ \bibinfo {author} {\bibfnamefont {D.~J.}\ \bibnamefont {Singh}},\ }\href {\doibase 10.1103/PhysRevB.78.085104} {\bibfield  {journal} {\bibinfo  {journal} {Phys. Rev. B}\ }\textbf {\bibinfo {volume} {78}},\ \bibinfo {pages} {085104} (\bibinfo {year} {2008})}\BibitemShut {NoStop}%
\bibitem [{\citenamefont {Ye}\ \emph {et~al.}(2014)\citenamefont {Ye}, \citenamefont {Bao}, \citenamefont {Chi}, \citenamefont {Fang}, \citenamefont {Wang}, \citenamefont {Mao}, \citenamefont {Wang}, \citenamefont {Liu}, \citenamefont {Jie-Ming}, \citenamefont {Santos},\ and\ \citenamefont {Molaison}}]{Ye_2014}%
  \BibitemOpen
  \bibfield  {author} {\bibinfo {author} {\bibfnamefont {F.}~\bibnamefont {Ye}}, \bibinfo {author} {\bibfnamefont {W.}~\bibnamefont {Bao}}, \bibinfo {author} {\bibfnamefont {S.-X.}\ \bibnamefont {Chi}}, \bibinfo {author} {\bibfnamefont {M.-H.}\ \bibnamefont {Fang}}, \bibinfo {author} {\bibfnamefont {H.-D.}\ \bibnamefont {Wang}}, \bibinfo {author} {\bibfnamefont {Q.-H.}\ \bibnamefont {Mao}}, \bibinfo {author} {\bibfnamefont {J.-C.}\ \bibnamefont {Wang}}, \bibinfo {author} {\bibfnamefont {J.-J.}\ \bibnamefont {Liu}}, \bibinfo {author} {\bibfnamefont {S.}~\bibnamefont {Jie-Ming}}, \bibinfo {author} {\bibfnamefont {A.~M.~d.}\ \bibnamefont {Santos}}, \ and\ \bibinfo {author} {\bibfnamefont {J.~J.}\ \bibnamefont {Molaison}},\ }\href {\doibase 10.1088/0256-307X/31/12/127401} {\bibfield  {journal} {\bibinfo  {journal} {Chinese Physics Letters}\ }\textbf {\bibinfo {volume} {31}},\ \bibinfo {pages} {127401} (\bibinfo {year} {2014})}\BibitemShut {NoStop}%
\bibitem [{\citenamefont {Tsuchiya}\ \emph {et~al.}(2017)\citenamefont {Tsuchiya}, \citenamefont {Ikeda}, \citenamefont {Zhang}, \citenamefont {Kishimoto}, \citenamefont {Kikegawa}, \citenamefont {Hirao}, \citenamefont {Kawaguchi}, \citenamefont {Ohishi},\ and\ \citenamefont {Kobayashi}}]{JPSJ.86.033705}%
  \BibitemOpen
  \bibfield  {author} {\bibinfo {author} {\bibfnamefont {Y.}~\bibnamefont {Tsuchiya}}, \bibinfo {author} {\bibfnamefont {S.}~\bibnamefont {Ikeda}}, \bibinfo {author} {\bibfnamefont {X.-W.}\ \bibnamefont {Zhang}}, \bibinfo {author} {\bibfnamefont {S.}~\bibnamefont {Kishimoto}}, \bibinfo {author} {\bibfnamefont {T.}~\bibnamefont {Kikegawa}}, \bibinfo {author} {\bibfnamefont {N.}~\bibnamefont {Hirao}}, \bibinfo {author} {\bibfnamefont {S.~I.}\ \bibnamefont {Kawaguchi}}, \bibinfo {author} {\bibfnamefont {Y.}~\bibnamefont {Ohishi}}, \ and\ \bibinfo {author} {\bibfnamefont {H.}~\bibnamefont {Kobayashi}},\ }\href {\doibase 10.7566/JPSJ.86.033705} {\bibfield  {journal} {\bibinfo  {journal} {Journal of the Physical Society of Japan}\ }\textbf {\bibinfo {volume} {86}},\ \bibinfo {pages} {033705} (\bibinfo {year} {2017})},\ \Eprint {http://arxiv.org/abs/https://doi.org/10.7566/JPSJ.86.033705} {https://doi.org/10.7566/JPSJ.86.033705} \BibitemShut {NoStop}%
\bibitem [{\citenamefont {Bendele}\ \emph {et~al.}(2013)\citenamefont {Bendele}, \citenamefont {Marini}, \citenamefont {Joseph}, \citenamefont {Pierantozzi}, \citenamefont {Caporale}, \citenamefont {Bianconi}, \citenamefont {Pomjakushina}, \citenamefont {Conder}, \citenamefont {Krzton-Maziopa}, \citenamefont {Irifune}, \citenamefont {Shinmei}, \citenamefont {Pascarelli}, \citenamefont {Dore}, \citenamefont {Saini},\ and\ \citenamefont {Postorino}}]{PhysRevB.88.180506}%
  \BibitemOpen
  \bibfield  {author} {\bibinfo {author} {\bibfnamefont {M.}~\bibnamefont {Bendele}}, \bibinfo {author} {\bibfnamefont {C.}~\bibnamefont {Marini}}, \bibinfo {author} {\bibfnamefont {B.}~\bibnamefont {Joseph}}, \bibinfo {author} {\bibfnamefont {G.~M.}\ \bibnamefont {Pierantozzi}}, \bibinfo {author} {\bibfnamefont {A.~S.}\ \bibnamefont {Caporale}}, \bibinfo {author} {\bibfnamefont {A.}~\bibnamefont {Bianconi}}, \bibinfo {author} {\bibfnamefont {E.}~\bibnamefont {Pomjakushina}}, \bibinfo {author} {\bibfnamefont {K.}~\bibnamefont {Conder}}, \bibinfo {author} {\bibfnamefont {A.}~\bibnamefont {Krzton-Maziopa}}, \bibinfo {author} {\bibfnamefont {T.}~\bibnamefont {Irifune}}, \bibinfo {author} {\bibfnamefont {T.}~\bibnamefont {Shinmei}}, \bibinfo {author} {\bibfnamefont {S.}~\bibnamefont {Pascarelli}}, \bibinfo {author} {\bibfnamefont {P.}~\bibnamefont {Dore}}, \bibinfo {author} {\bibfnamefont {N.~L.}\ \bibnamefont {Saini}}, \ and\ \bibinfo {author} {\bibfnamefont {P.}~\bibnamefont {Postorino}},\ }\href {\doibase
  10.1103/PhysRevB.88.180506} {\bibfield  {journal} {\bibinfo  {journal} {Phys. Rev. B}\ }\textbf {\bibinfo {volume} {88}},\ \bibinfo {pages} {180506} (\bibinfo {year} {2013})}\BibitemShut {NoStop}%
\bibitem [{\citenamefont {Yamamoto}\ \emph {et~al.}(2016)\citenamefont {Yamamoto}, \citenamefont {Yamaoka}, \citenamefont {Tanaka}, \citenamefont {Okazaki}, \citenamefont {Ozaki}, \citenamefont {Takano}, \citenamefont {Lin}, \citenamefont {Fujita}, \citenamefont {Kagayama}, \citenamefont {Shimizu}, \citenamefont {Hiraoka}, \citenamefont {Ishii}, \citenamefont {Liao}, \citenamefont {Tsuei},\ and\ \citenamefont {Mizuki}}]{Yamamoto:2016aa}%
  \BibitemOpen
  \bibfield  {author} {\bibinfo {author} {\bibfnamefont {Y.}~\bibnamefont {Yamamoto}}, \bibinfo {author} {\bibfnamefont {H.}~\bibnamefont {Yamaoka}}, \bibinfo {author} {\bibfnamefont {M.}~\bibnamefont {Tanaka}}, \bibinfo {author} {\bibfnamefont {H.}~\bibnamefont {Okazaki}}, \bibinfo {author} {\bibfnamefont {T.}~\bibnamefont {Ozaki}}, \bibinfo {author} {\bibfnamefont {Y.}~\bibnamefont {Takano}}, \bibinfo {author} {\bibfnamefont {J.-F.}\ \bibnamefont {Lin}}, \bibinfo {author} {\bibfnamefont {H.}~\bibnamefont {Fujita}}, \bibinfo {author} {\bibfnamefont {T.}~\bibnamefont {Kagayama}}, \bibinfo {author} {\bibfnamefont {K.}~\bibnamefont {Shimizu}}, \bibinfo {author} {\bibfnamefont {N.}~\bibnamefont {Hiraoka}}, \bibinfo {author} {\bibfnamefont {H.}~\bibnamefont {Ishii}}, \bibinfo {author} {\bibfnamefont {Y.-F.}\ \bibnamefont {Liao}}, \bibinfo {author} {\bibfnamefont {K.-D.}\ \bibnamefont {Tsuei}}, \ and\ \bibinfo {author} {\bibfnamefont {J.}~\bibnamefont {Mizuki}},\ }\href {\doibase 10.1038/srep30946} {\bibfield
  {journal} {\bibinfo  {journal} {Scientific Reports}\ }\textbf {\bibinfo {volume} {6}},\ \bibinfo {pages} {30946} (\bibinfo {year} {2016})}\BibitemShut {NoStop}%
\bibitem [{\citenamefont {Yu}\ \emph {et~al.}(2011)\citenamefont {Yu}, \citenamefont {Goswami},\ and\ \citenamefont {Si}}]{PhysRevB.84.094451}%
  \BibitemOpen
  \bibfield  {author} {\bibinfo {author} {\bibfnamefont {R.}~\bibnamefont {Yu}}, \bibinfo {author} {\bibfnamefont {P.}~\bibnamefont {Goswami}}, \ and\ \bibinfo {author} {\bibfnamefont {Q.}~\bibnamefont {Si}},\ }\href {\doibase 10.1103/PhysRevB.84.094451} {\bibfield  {journal} {\bibinfo  {journal} {Phys. Rev. B}\ }\textbf {\bibinfo {volume} {84}},\ \bibinfo {pages} {094451} (\bibinfo {year} {2011})}\BibitemShut {NoStop}%
\bibitem [{\citenamefont {Jaeschke-Ubiergo}\ \emph {et~al.}(2024)\citenamefont {Jaeschke-Ubiergo}, \citenamefont {Bharadwaj}, \citenamefont {Jungwirth}, \citenamefont {\ifmmode~\check{S}\else \v{S}\fi{}mejkal},\ and\ \citenamefont {Sinova}}]{PhysRevB.109.094425}%
  \BibitemOpen
  \bibfield  {author} {\bibinfo {author} {\bibfnamefont {R.}~\bibnamefont {Jaeschke-Ubiergo}}, \bibinfo {author} {\bibfnamefont {V.~K.}\ \bibnamefont {Bharadwaj}}, \bibinfo {author} {\bibfnamefont {T.}~\bibnamefont {Jungwirth}}, \bibinfo {author} {\bibfnamefont {L.}~\bibnamefont {\ifmmode~\check{S}\else \v{S}\fi{}mejkal}}, \ and\ \bibinfo {author} {\bibfnamefont {J.}~\bibnamefont {Sinova}},\ }\href {\doibase 10.1103/PhysRevB.109.094425} {\bibfield  {journal} {\bibinfo  {journal} {Phys. Rev. B}\ }\textbf {\bibinfo {volume} {109}},\ \bibinfo {pages} {094425} (\bibinfo {year} {2024})}\BibitemShut {NoStop}%
\bibitem [{\citenamefont {Svitlyk}\ \emph {et~al.}(2011)\citenamefont {Svitlyk}, \citenamefont {Chernyshov}, \citenamefont {Pomjakushina}, \citenamefont {Krzton-Maziopa}, \citenamefont {Conder}, \citenamefont {Pomjakushin},\ and\ \citenamefont {Dmitriev}}]{ic201160y}%
  \BibitemOpen
  \bibfield  {author} {\bibinfo {author} {\bibfnamefont {V.}~\bibnamefont {Svitlyk}}, \bibinfo {author} {\bibfnamefont {D.}~\bibnamefont {Chernyshov}}, \bibinfo {author} {\bibfnamefont {E.}~\bibnamefont {Pomjakushina}}, \bibinfo {author} {\bibfnamefont {A.}~\bibnamefont {Krzton-Maziopa}}, \bibinfo {author} {\bibfnamefont {K.}~\bibnamefont {Conder}}, \bibinfo {author} {\bibfnamefont {V.}~\bibnamefont {Pomjakushin}}, \ and\ \bibinfo {author} {\bibfnamefont {V.}~\bibnamefont {Dmitriev}},\ }\href {\doibase 10.1021/ic201160y} {\bibfield  {journal} {\bibinfo  {journal} {Inorganic Chemistry}\ }\textbf {\bibinfo {volume} {50}},\ \bibinfo {pages} {10703} (\bibinfo {year} {2011})},\ \bibinfo {note} {pMID: 21988233},\ \Eprint {http://arxiv.org/abs/https://doi.org/10.1021/ic201160y} {https://doi.org/10.1021/ic201160y} \BibitemShut {NoStop}%
\bibitem [{\citenamefont {Leraand}\ \emph {et~al.}(2025)\citenamefont {Leraand}, \citenamefont {M{\ae}land},\ and\ \citenamefont {Sudb{\o}}}]{leraand2025phononmediatedspinpolarizedsuperconductivityaltermagnets}%
  \BibitemOpen
  \bibfield  {author} {\bibinfo {author} {\bibfnamefont {K.}~\bibnamefont {Leraand}}, \bibinfo {author} {\bibfnamefont {K.}~\bibnamefont {M{\ae}land}}, \ and\ \bibinfo {author} {\bibfnamefont {A.}~\bibnamefont {Sudb{\o}}},\ }\href {https://arxiv.org/abs/2502.08704} {\enquote {\bibinfo {title} {Phonon-mediated spin-polarized superconductivity in altermagnets},}\ } (\bibinfo {year} {2025}),\ \Eprint {http://arxiv.org/abs/2502.08704} {arXiv:2502.08704 [cond-mat.supr-con]} \BibitemShut {NoStop}%
\end{thebibliography}%


\begin{thebibliography}{8}
\expandafter\ifx\csname natexlab\endcsname\relax\def\natexlab#1{#1}\fi
\expandafter\ifx\csname bibnamefont\endcsname\relax
  \def\bibnamefont#1{#1}\fi
\expandafter\ifx\csname bibfnamefont\endcsname\relax
  \def\bibfnamefont#1{#1}\fi
\expandafter\ifx\csname citenamefont\endcsname\relax
  \def\citenamefont#1{#1}\fi
\expandafter\ifx\csname url\endcsname\relax
  \def\url#1{\texttt{#1}}\fi
\expandafter\ifx\csname urlprefix\endcsname\relax\def\urlprefix{URL }\fi
\providecommand{\bibinfo}[2]{#2}
\providecommand{\eprint}[2][]{\url{#2}}

\bibitem[{\citenamefont{Kresse and Joubert}(1999)}]{method:pawvasp}
\bibinfo{author}{\bibfnamefont{G.}~\bibnamefont{Kresse}} \bibnamefont{and} \bibinfo{author}{\bibfnamefont{D.}~\bibnamefont{Joubert}}, \bibinfo{journal}{Phys. Rev. B} \textbf{\bibinfo{volume}{59}}, \bibinfo{pages}{1758} (\bibinfo{year}{1999}).

\bibitem[{\citenamefont{Perdew et~al.}(1996)\citenamefont{Perdew, Burke, and Ernzerhof}}]{method:pbe}
\bibinfo{author}{\bibfnamefont{J.~P.} \bibnamefont{Perdew}}, \bibinfo{author}{\bibfnamefont{K.}~\bibnamefont{Burke}}, \bibnamefont{and} \bibinfo{author}{\bibfnamefont{M.}~\bibnamefont{Ernzerhof}}, \bibinfo{journal}{Phys. Rev. Lett.} \textbf{\bibinfo{volume}{77}}, \bibinfo{pages}{3865} (\bibinfo{year}{1996}).

\bibitem[{\citenamefont{Gao et~al.}(2017)\citenamefont{Gao, Kong, Yan, Lu, and Xiang}}]{PhysRevB.95.174523}
\bibinfo{author}{\bibfnamefont{M.}~\bibnamefont{Gao}}, \bibinfo{author}{\bibfnamefont{X.}~\bibnamefont{Kong}}, \bibinfo{author}{\bibfnamefont{X.-W.} \bibnamefont{Yan}}, \bibinfo{author}{\bibfnamefont{Z.-Y.} \bibnamefont{Lu}}, \bibnamefont{and} \bibinfo{author}{\bibfnamefont{T.}~\bibnamefont{Xiang}}, \bibinfo{journal}{Phys. Rev. B} \textbf{\bibinfo{volume}{95}}, \bibinfo{pages}{174523} (\bibinfo{year}{2017}), \urlprefix\url{https://link.aps.org/doi/10.1103/PhysRevB.95.174523}.

\bibitem[{\citenamefont{Xu et~al.}(2025)\citenamefont{Xu, Wu, Zhi, Cao, Dai, Cao, Wang, and Lin}}]{Xu:2025aa}
\bibinfo{author}{\bibfnamefont{C.}~\bibnamefont{Xu}}, \bibinfo{author}{\bibfnamefont{S.}~\bibnamefont{Wu}}, \bibinfo{author}{\bibfnamefont{G.-X.} \bibnamefont{Zhi}}, \bibinfo{author}{\bibfnamefont{G.}~\bibnamefont{Cao}}, \bibinfo{author}{\bibfnamefont{J.}~\bibnamefont{Dai}}, \bibinfo{author}{\bibfnamefont{C.}~\bibnamefont{Cao}}, \bibinfo{author}{\bibfnamefont{X.}~\bibnamefont{Wang}}, \bibnamefont{and} \bibinfo{author}{\bibfnamefont{H.-Q.} \bibnamefont{Lin}}, \bibinfo{journal}{Nature Communications} \textbf{\bibinfo{volume}{16}}, \bibinfo{pages}{3114} (\bibinfo{year}{2025}), \urlprefix\url{https://doi.org/10.1038/s41467-025-58446-6}.

\bibitem[{\citenamefont{Mazin et~al.}(2008)\citenamefont{Mazin, Johannes, Boeri, Koepernik, and Singh}}]{PhysRevB.78.085104}
\bibinfo{author}{\bibfnamefont{I.~I.} \bibnamefont{Mazin}}, \bibinfo{author}{\bibfnamefont{M.~D.} \bibnamefont{Johannes}}, \bibinfo{author}{\bibfnamefont{L.}~\bibnamefont{Boeri}}, \bibinfo{author}{\bibfnamefont{K.}~\bibnamefont{Koepernik}}, \bibnamefont{and} \bibinfo{author}{\bibfnamefont{D.~J.} \bibnamefont{Singh}}, \bibinfo{journal}{Phys. Rev. B} \textbf{\bibinfo{volume}{78}}, \bibinfo{pages}{085104} (\bibinfo{year}{2008}), \urlprefix\url{https://link.aps.org/doi/10.1103/PhysRevB.78.085104}.

\bibitem[{\citenamefont{Schwarz et~al.}(2002)\citenamefont{Schwarz, Blaha, and Madsen}}]{method:wien2k}
\bibinfo{author}{\bibfnamefont{K.}~\bibnamefont{Schwarz}}, \bibinfo{author}{\bibfnamefont{P.}~\bibnamefont{Blaha}}, \bibnamefont{and} \bibinfo{author}{\bibfnamefont{G.~K.~H.} \bibnamefont{Madsen}}, \bibinfo{journal}{Computer Physics Communications} \textbf{\bibinfo{volume}{147}}, \bibinfo{pages}{71} (\bibinfo{year}{2002}).

\bibitem[{\citenamefont{Kotliar et~al.}(2006)\citenamefont{Kotliar, Savrasov, Haule, Oudovenko, Parcollet, and Marianetti}}]{method:dmft2}
\bibinfo{author}{\bibfnamefont{G.}~\bibnamefont{Kotliar}}, \bibinfo{author}{\bibfnamefont{S.~Y.} \bibnamefont{Savrasov}}, \bibinfo{author}{\bibfnamefont{K.}~\bibnamefont{Haule}}, \bibinfo{author}{\bibfnamefont{V.~S.} \bibnamefont{Oudovenko}}, \bibinfo{author}{\bibfnamefont{O.}~\bibnamefont{Parcollet}}, \bibnamefont{and} \bibinfo{author}{\bibfnamefont{C.~A.} \bibnamefont{Marianetti}}, \bibinfo{journal}{Rev. Mod. Phys.} \textbf{\bibinfo{volume}{78}}, \bibinfo{pages}{865} (\bibinfo{year}{2006}).

\bibitem[{\citenamefont{Haule}(2007)}]{method:ctqmc_dmft}
\bibinfo{author}{\bibfnamefont{K.}~\bibnamefont{Haule}}, \bibinfo{journal}{Phys. Rev. B} \textbf{\bibinfo{volume}{75}}, \bibinfo{pages}{155113} (\bibinfo{year}{2007}).

\end{thebibliography}

\end{document}


\title{Supplementary Information: Pressure Induced Altermagnetism in Layered Ternary Iron-Selenides} \preprint{1}

\author{Zilong Li}
\author{Xin Ma}
 \affiliation{Center for Correlated Matter and School of Physics, Zhejiang University, Hangzhou 310058, China}

\author{Siqi Wu}
 \affiliation{Department of Physics, The Hong Kong University of Science and Technology, Clear Water Bay, Kowloon, Hong Kong, China}

\author{H.-Q. Yuan}
 \affiliation{Center for Correlated Matter and School of Physics, Zhejiang University, Hangzhou 310058, China}
 \affiliation{Institute for Advanced Study in Physics, Zhejiang University, Hangzhou 310058, China}

\author{Jianhui Dai}
\email[E-mail address: ]{daijh@hznu.edu.cn}
 \affiliation{School of Physics, Hangzhou Normal University, Hangzhou 310036, China}
 \affiliation{Institute for Advanced Study in Physics, Zhejiang University, Hangzhou 310058, China}

\author{Chao Cao}
 \email[E-mail address: ]{ccao@zju.edu.cn}
 \affiliation{Center for Correlated Matter and School of Physics, Zhejiang University, Hangzhou 310058, China}
 \affiliation{Institute for Advanced Study in Physics, Zhejiang University, Hangzhou 310058, China}

\date{Feb. 21, 2025}

\maketitle

\subsection{DFT Calculation Details}
\begin{figure}[h]
 \includegraphics[width=5cm]{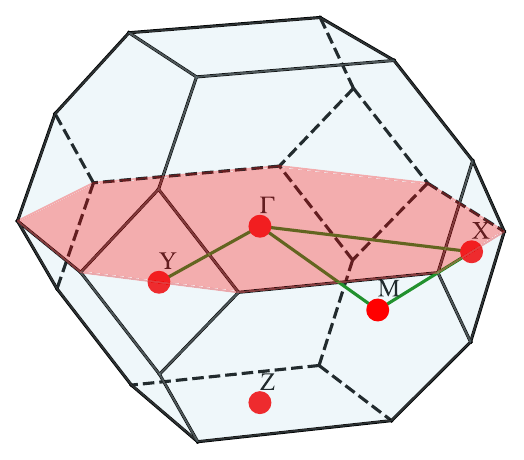}
 \caption{The Brillouin zone and high symmetry points.\label{fig:BZ}}
\end{figure}

The density functional calculations are performed using projected augmented wave (PAW) method as implemented in VASP\cite{method:pawvasp}. The Perdew, Burke, Ernzerhof parameterization of generalized gradient approximation (PBE)\cite{method:pbe} is employed as the exchange-correlation functional. Plane-wave basis up to 450 eV and $\Gamma$-centered K-mesh with k-spacing less than 0.3 \AA$^{-1}$ is employed to ensure convergency of total energy to order of 1 meV/atom. The Brillouin zone and the high symmetry points of both BS-AFM and \neelfm\ phase (space group $I$4/m, No. 87) are indicated in the FIG. \ref{fig:BZ}.

\begin{figure}[h]
 \includegraphics[width=14cm]{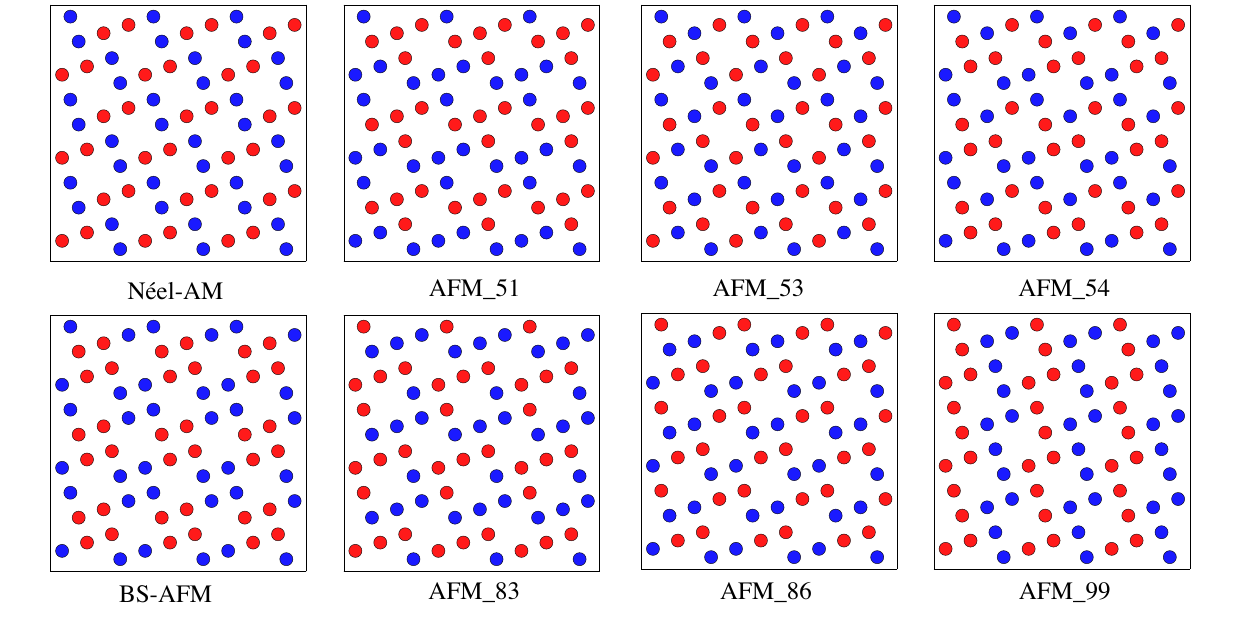}
 \caption{Magnetic configurations in size-1 cell. Only Fe-sites are shown, and up/down atoms are indicated by red/blue color. The AFM-53 phase is the checker-board AFM phase, and the AFM-86 phase is dubbed as the ``pair-checkerboard AFM" in Ref. \onlinecite{PhysRevB.95.174523}.\label{fig:magconf1}}
\end{figure}
\subsection{Magnetic Configurations in Size-1 Cell}
The high-throughput search of the magnetic configurations is performed using the same algorithm as described in Ref. \onlinecite{Xu:2025aa}. For size-1 cell, the search yields 8 possible configurations, as shown in FIG. \ref{fig:magconf1}.

\subsection{Stability of \neelfm\ Phase}
\begin{figure}[h]
  \includegraphics[width=16cm]{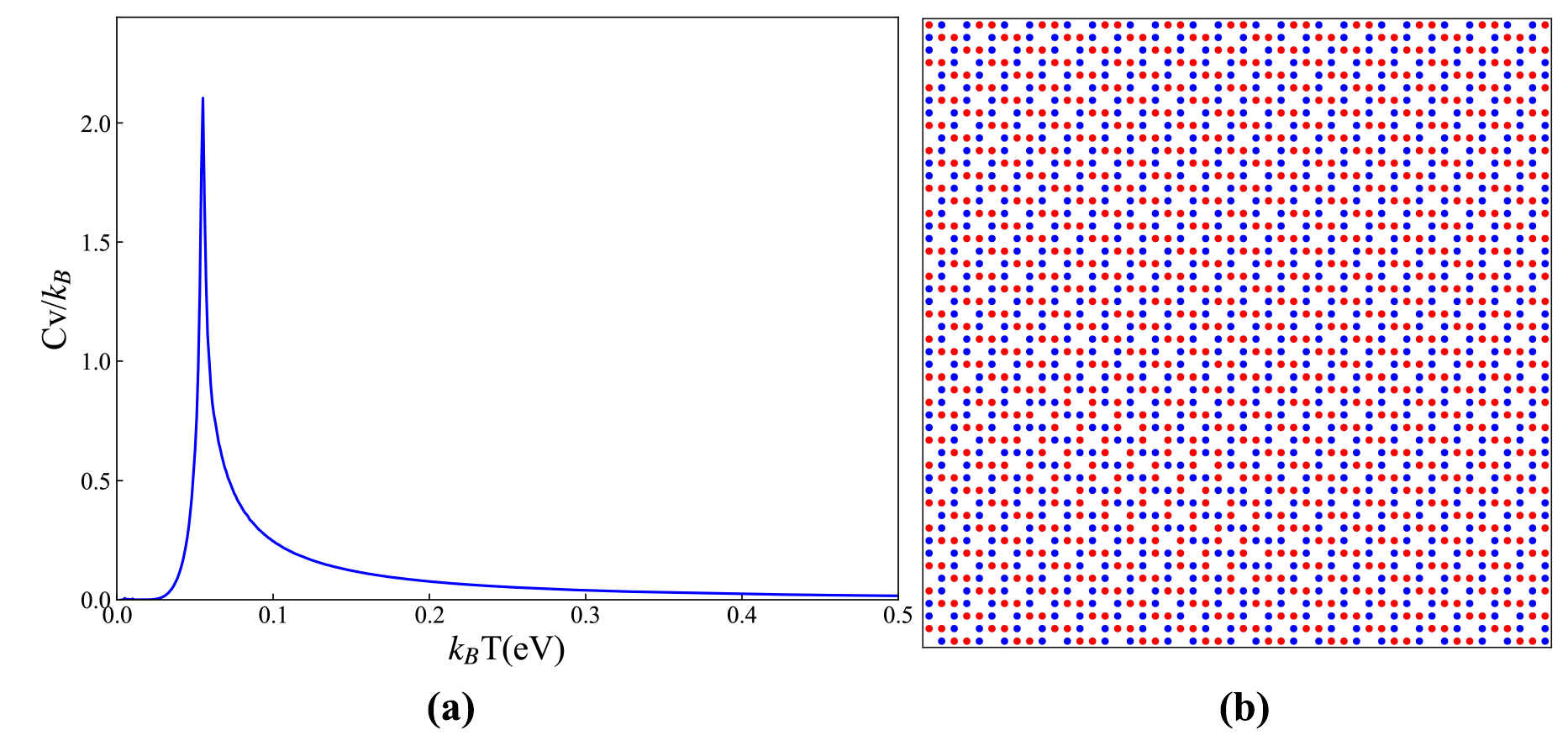}
 \caption{ (a) Specific heat ($C_v(T)$) and (b) Equilibrium configuration at $k_BT$=15 meV obtained from classical Monte Carlo simulation of extended $J_1$-$J_2$ model using \rb245\ parameters at 12 GPa. Red/blue dots indicate spin up/down sites in (b). \label{fig:mc}}
\end{figure}

We also verify the stability of the \neelfm\ phase using classical Monte Carlo simulation assuming extended $J_1$-$J_2$ model. The employed parameters was obtained by fitting DFT parameters of size-1 results of \rb245\ at 12 GPa, i.e. $J_1=50.1$ meV$/S^2$, $J'_1$=37.6 meV$/S^2$, $J_2$=11.9 meV$/S^2$, $J'_2$=44.4 meV$/S^2$. The specific heat (Fig. \ref{fig:mc}a) indicates a phase transition at $k_BT\approx$55 meV, and the equilibrium state at $k_B T$=15 meV is the \neelfm\ phase as shown in FIG. \ref{fig:mc}(b).

\begin{figure}[h]
  \includegraphics[width=8cm]{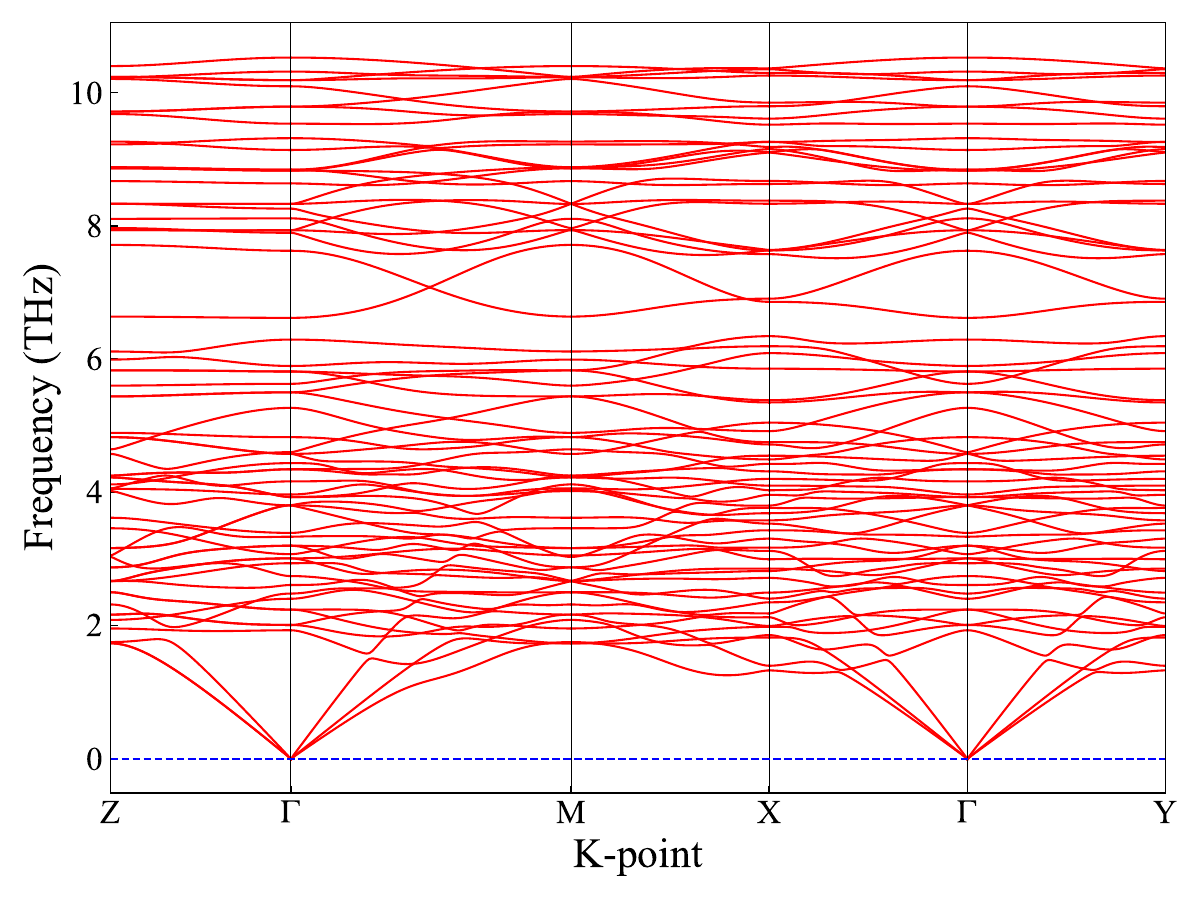}
 \caption{Phonon spectrum of Rb$_2$Fe$_4$Se$_5$ in the \neelfm\ phase at 12 GPa. \label{fig:phonon}}
\end{figure}

The structural dynamic stability of the \neelfm\ phase is also verified by calculating the phonon spectrum at 12 GPa (FIG. \ref{fig:phonon}).

\subsection{Magnetic Configurations in Size-2 Cell}
\begin{figure}[h]
 \includegraphics[width=14cm]{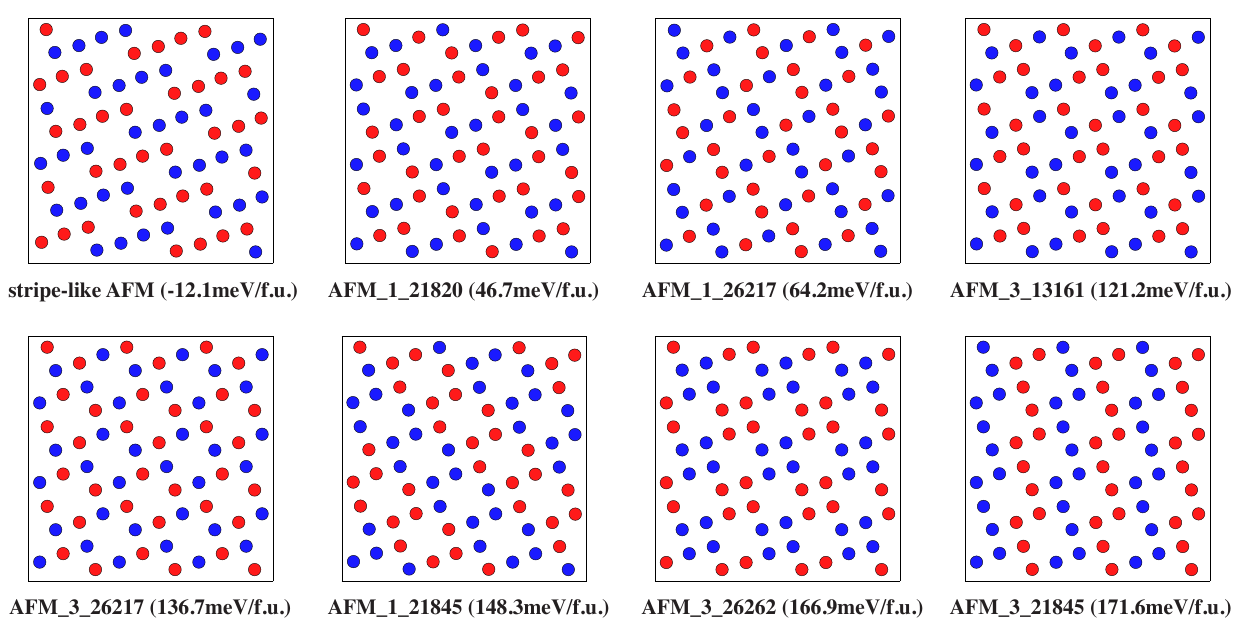}
 \caption{Stable magnetic configurations of Rb$_2$Fe$_4$Se$_5$ in size-2 cell at 12 GPa. Only Fe-sites are shown, and up/down atoms are indicated by red/blue color. The numbers in the brackets are the enthalpies relative to the \neelfm\ phase.\label{fig:magconf2}}
\end{figure}

Under high pressure, most of the size-2 phases are unstable, and become either nonmagnetic or other AFM phases. In FIG. \ref{fig:magconf2}, we show the stable size-2 configurations and their relative enthalpies (in unit of meV/f.u.) to the \neelfm\ phase at 12 GPa.

\subsection{DFT+DMFT Calculations at 12 GPa}

Since DFT tends to overestimate the magnetic interactions in the iron-based superconductors\citep{PhysRevB.78.085104}, we also calculate the total energies of the BS-AFM, \neelfm, and stripe-like AFM phases at 12 GPa to verify the results. Due to the large amount of computation involved in DMFT calculations, it is not practical to perform full structural relaxation. Therefore, we take the structures fully optimized under DFT to perform these calculations. The calculations were performed using Wien2K+EDMFTF package\cite{method:wien2k,method:dmft2,method:ctqmc_dmft}. Muffin-tin radii of Rb, Fe, and Se atoms were set to 2.50, 2.08 and 1.98 \AA, and $RK_{\mathrm{max}}=7$ in all calculations. The hybridization-expansion continuous-time quantum monte-carlo (CTQMC) impurity solver was employed for Fe-3d orbitals with $U=5.0$ eV and $J=0.8$ eV. We considered two cases for each phase, 116 K with nominal double-counting scheme fixed at $n_d=6.0$ and 290 K with exact double-counting scheme. The \neelfm\ phase has lower enthalpy than the stripe-like AFM phase in both calculations. In addition, we see that the ordered moments in DFT+DMFT calculations are more reasonable than those from simple DFT calculations, and agrees better with experiments. 
\begin{table}[h]
 \caption{DFT+DMFT total enthalpies ($H=E+pV$) and staggered moments (numbers inside the brackets) of the BS-AFM, \neelfm, and stripe-like AFM phases \rb245\ at 12 GPa. Case 1 is 116 K calculation with nominal double-counting (DC) fixed at $n_d=6.0$; case 2 is 290 K calculation with exact DC. All enthalpies are relative to BS-AFM and in unit of meV/f.u., and moments are in $\mu_B$/Fe. For stripe-like AFM phase, there are two inequivalent Fe sites, and the staggered moments for both sites are given.\label{tab:dmft}}
 \begin{tabular}{c|c|c||c|c|c}
 \multicolumn{3}{c||}{116K nominal DC} & \multicolumn{3}{c}{290K exact DC} \\
 \neelfm\ & BS-AFM & stripe-like AFM & \neelfm\ & BS-AFM & stripe-like AFM \\
 \hline\hline
 -708.1 (0.8)  &  0.0 (2.4)   &   -601.7 (1.2, 1.0)   & -266.2 (1.4)  &  0.0 (2.6)   &   -117.4 (1.8, 1.4)   \\
 \end{tabular}
 \end{table}

\begin{figure}[h]
 \includegraphics[width=16cm]{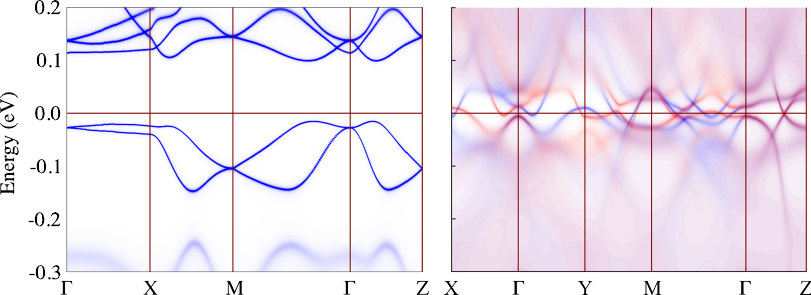}
 \caption{Momentum dependent spectral function $A(\mathbf{k}, \omega)$ of (a) BS-AFM and (b) \neelfm\ phase \rb245\ under 12 GPa at 116K from DFT+DMFT calculations. Contributions from different spin channels are indicated with red/blue color in (b). \label{fig:dmftbs}}
\end{figure}

We also demonstrate the many-body spectral function from DFT+DMFT calculations in FIG. \ref{fig:dmftbs}. The BS-AFM phase remains gapped with $E_g\approx120$ meV in DFT+DMFT calculations. The spin-splitting $\Delta_s$ of the \neelfm\ phase is also evident. Both $E_g$ and $\Delta_s$ are moderately renormalized compared to DFT results.

\subsection{Results of $\mathrm{K}_2\mathrm{Fe}_4\mathrm{Se}_5$, $\mathrm{Na}_2\mathrm{Fe}_4\mathrm{Se}_5$, $\mathrm{Tl}_2\mathrm{Fe}_4\mathrm{Se}_5$ and $\mathrm{Cs}_2\mathrm{Fe}_4\mathrm{Se}_5$}
\begin{figure}[h]
 \includegraphics[width=16cm]{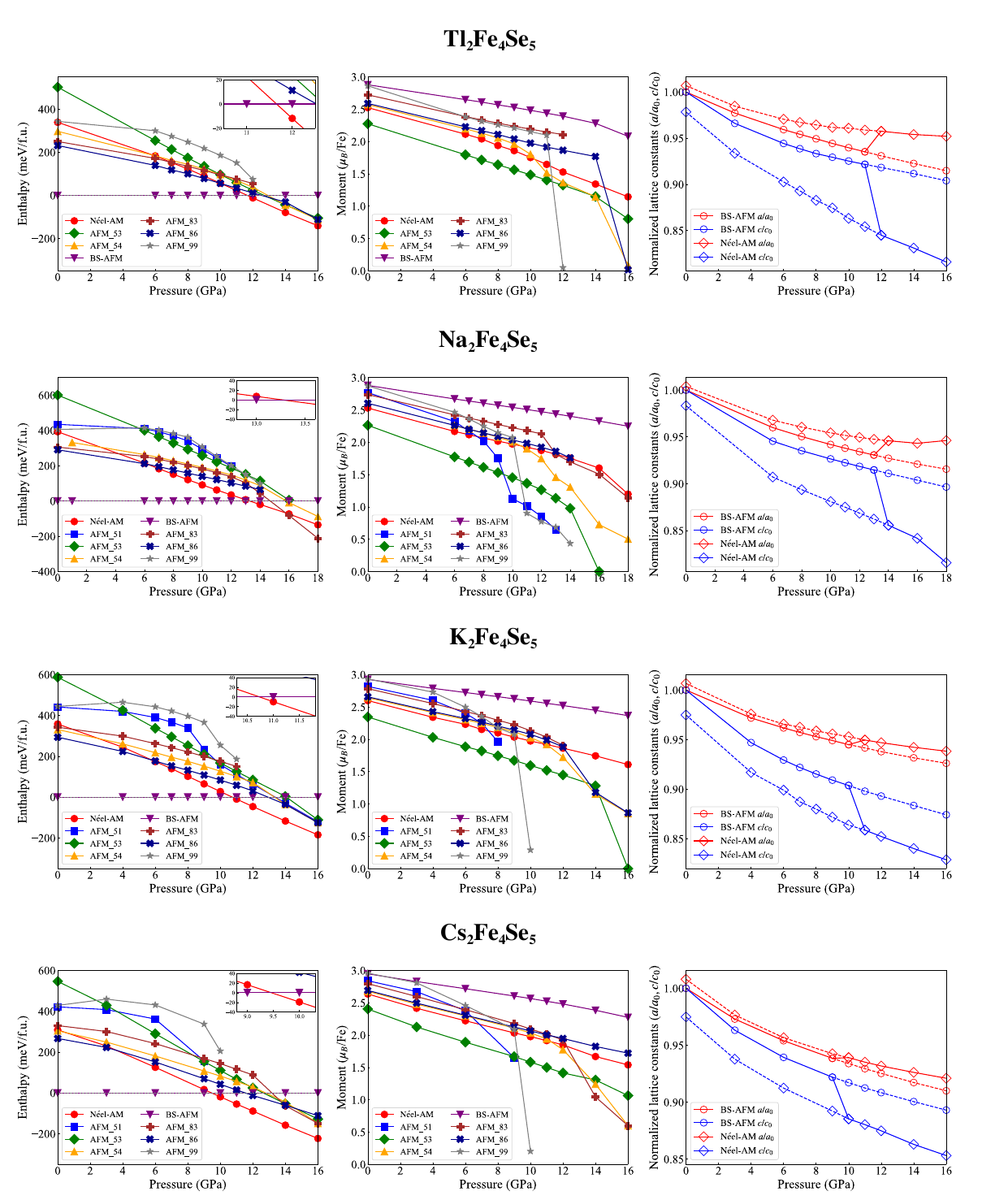}
 \caption{The total enthalpy, staggered moments and the normalized lattice constants of \tl245, \na245, \k245, \cs245. \label{fig:other245}}
\end{figure}

We show the evolution of enthalpy, staggered moment and normalized lattice constants of \tl245, \na245, \k245, and \cs245\ with respect to pressure in FIG. \ref{fig:other245}. In general, we see that there is always a BS-AFM to \neelfm\ phase transition under high pressure in these compounds. In addition, the variation of $\Delta c/c_0$ is always much larger than $\Delta a/a_0$ at the transition.

\subsection{Extended $J_1-J_2$ Model Fitting}
\begin{table}[htp]
 \caption{Total energies (meV/f.u.) and fitted $J_1$, $J'_1$, $J_2$, $J'_2$ and $R^2$ of \rb245\ under different pressure/$c$-strain (positive/negative values in the first column). All energies are respective to the BS-AFM phase. \label{tab:fitting}}
 \begin{tabular}{c||c|c|c|c|c|c|c||c|c|c|c|c}
 Pressure                & \multicolumn{7}{|c||}{Phase Energies (meV/f.u.)}                       & \multicolumn{4}{c|}{Parameters (meV/$S^2$)} &  \\
  Strain & BS-AFM & \neelfm\ & AFM-53 & AFM-54 & AFM-83 & AFM-86 & AFM-99 & $J_1$ & $J'_1$ & $J_2$ & $J'_2$ & $R^2$ \\
 \hline\hline
   0 GPa         & 0.0    & 1222.2    & 1735.0       &  861.4        & 1119.5       & 822.5       & 1397.4       &  -40.7     &  71.0      &  4.9     &  67.7      &  0.9999     \\
   3 GPa         & 0.0    & 1087.8    & 1525.9       & 782.3         & 1085.3       & 727.0       &  1437.1      &  -24.1     &  86.9      & 3.6      &  71.2      &  0.9994     \\
   6 GPa         & 0.0    & 904.4     & 1294.9       &  675.4        & 1003.6       & 611.3       &  1387.9      &  -8.6     &   94.8     &  3.5     &  72.3      &  0.9988     \\
   9 GPa         & 0.0    & 705.7     & 1064.8       &  559.2        &  891.3       & 490.5       &  1246.5      &  2.4     &   92.2     &  3.4     &   69.0     &   0.9982    \\
  10 GPa         & 0.0    & 638.7     & 989.8        &  519.7        &  848.8       &  450.6      &  1189.1      &   5.5    &   90.1     & 3.5      &  67.4      &  0.9979     \\
 \hline
  -1\%           & 0.0    & 541.3     & 855.4       & 445.3          & 780.6       & 378.3       &  1120.8      &  11.9     &   90.8     &  4.1     & 65.4       &  0.9979     \\
  -3\%           & 0.0    & 334.9     & 577.3       & 287.6          & 618.4       & 225.2       &  950.5      &   23.9    &   89.2     &  6.0     &  60.0      &  0.9981     \\
  -5\%           & 0.0    & 115.7     & 290.7       & 119.6          & 434.0       & 60.8       &   693.8     &  32.1     &   78.5     &  7.9     &  50.2      &  0.9940     \\
  -7\%           & 0.0    & -112.9    & 6.3       &  -55.1           & 143.3       &  -113.2      &  343.3      &  35.4     &   57.0     &  13.2     &  35.8      &  0.9965     \\
 \end{tabular}
 \end{table}
 To fit the exchange interactions in the extended $J_1-J_2$ model under different pressure, we employ the fully relaxed ground state crystal structure at the specific pressure, and statically calculate the total energies different magnetic configurations. This is preferred over fitting the energies or enthalpies of relaxed structures because the exchange interactions $J$ conceptually originate from electronic hoppings and interactions, and does not involve electron-ion interactions. The fitting is done using least-square procedure considering 7 magnetic states in size-1 cell. The AFM-51 phase is not considered for 2 reasons: firstly, the AFM-51 phase is trivial in the extended $J_1-J_2$ model because all the coefficients of $J$ parameters are 0 for the AFM-51 phase; and secondly, the AFM-51 phase is not stable under high pressure. We list the fitted $J$s and the respective $R^2$ in the TAB. \ref{tab:fitting}. 

\subsection{Band Structure of $\mathrm{Rb}_2\mathrm{Fe}_4\mathrm{Se}_5$ in the BS-AFM phase}
\begin{figure}[h]
 \includegraphics[width=16cm]{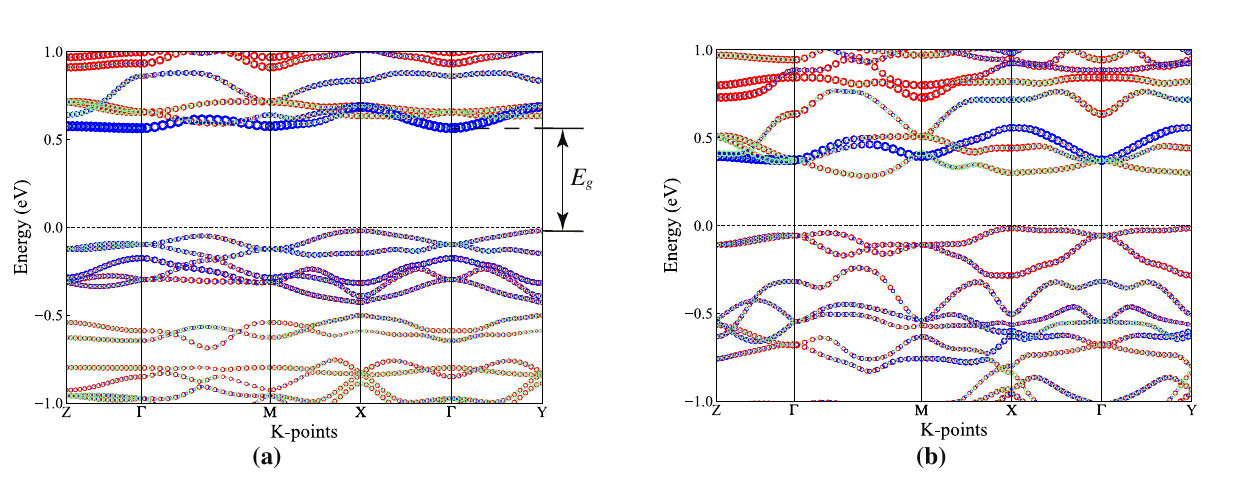}
\caption{The band structure of \rb245\ in BS-AFM phase under (a) ambient pressure and (b) 10 GPa. The sizes of the red, blue and light-green circles are proportional to the weight of Fe-3$d_{zx(y)}$, Fe-3$d_{z^2}$, and Fe-3$d_{x^2-y^2}$ orbitals, respectively.\label{fig:bsafm_bs}}
\end{figure}

We show the band structure of \rb245\ in BS-AFM phase under ambient pressure and 10 GPa in FIG. \ref{fig:bsafm_bs}. The indirect gap between $\Gamma$ and X(Y) is also indicated. The gap at ambient pressure is $E_g\approx 580 meV$. External pressure monotonically suppress it to 290 meV at 10 GPa. 

\subsection{RPA and Weak Coupling Analysis}
We focus on two possible cases. 1) Superconductivity (SC) emerges on the paramagnetic normal state by suppressing the \neelfm\ order; and 2) SC emerges on the \neelfm\ ordered normal state directly. They corresponds to scenario 2 and 3 in the main text. For the second case, a set of spin-polarized tight-binding Hamiltonians without SOC was obtained by fitting the first principles electronic structure of the \neelfm\ state \rb245\ at 14 GPa consists using maximally projected Wannier function method with all Fe-3d orbitals and Se-4p orbitals. For this spin-polarized case, we calculate the general bare electron susceptibility using:

$$[\chi^0(\mathbf{q}, i\nu)]_{st\sigma\sigma}^{pq\sigma\sigma}=-\frac{1}{N_{\mathbf{k}}\beta}\sum_{n,\mathbf{k}}G^{0\sigma}_{sp}(i\omega_n+i\nu, \mathbf{k+q})G^{0\sigma}_{qt}(i\omega_n, \mathbf{k})$$
$$[\chi^0(\mathbf{q}, i\nu)]_{st\sigma\bar{\sigma}}^{pq\sigma\bar{\sigma}}=-\frac{1}{N_{\mathbf{k}}\beta}\sum_{n,\mathbf{k}}G^{0\sigma}_{sp}(i\omega_n+i\nu, \mathbf{k+q})G^{0\bar{\sigma}}_{qt}(i\omega_n, \mathbf{k})$$
where $s,p,q,t$ are orbital indicies, $\sigma\in(\uparrow,\downarrow)$ is the spin index, $\bar{\sigma}$ is the opposite spin, $\omega_n=(2n+1)\pi/\beta$ are Matsubara frequencies, $\beta^{-1}=k_BT$, and $G^0$ is the non-interacting Green's function.

Considering on-site interactions on Fe-3d orbitals (we chose $U=$1.6 eV and $J=$0.1 eV), which takes the form:

$$V_{ii\sigma\sigma}^{ii\bar{\sigma}\bar{\sigma}}=-U, V_{ii\sigma\bar{\sigma}}^{ii\sigma\bar{\sigma}}=U, V_{ij\sigma\bar{\sigma}}^{ji\bar{\sigma}\sigma}=-J, V_{ij\sigma\bar{\sigma}}^{ji\sigma\bar{\sigma}}=J$$

$$V_{ii\sigma\sigma}^{jj\sigma\sigma}=-(U-3J), V_{ii\sigma\sigma}^{jj\bar{\sigma}\bar{\sigma}}=-(U-2J), V_{ii\sigma\bar{\sigma}}^{jj\sigma\bar{\sigma}}=J$$

$$V_{ij\sigma\sigma}^{ij\sigma\sigma}=(U-3J), V_{ij\sigma\sigma}^{ij\bar{\sigma}\bar{\sigma}}=-J, V_{ij\sigma\bar{\sigma}}^{ij\sigma\bar{\sigma}}=(U-2J)$$
  
the general susceptibility can be obtained using RPA:

\begin{eqnarray*}
\chi^{\uu}_{\uu}&=&\chi^{\uu}_{0\uu}+\chi^{\uu}_{0\uu}V^{\uu}_{\uu}\chi^{\uu}_{\uu}+\chi^{\uu}_{0\uu}V^{\dd}_{\uu}\chi^{\uu}_{\dd}+\chi^{\dd}_{0\uu}V^{\uu}_{\dd}\chi^{\uu}_{\uu}+\chi^{\dd}_{0\uu}V^{\dd}_{\dd}\chi^{\uu}_{\dd} \\
\chi^{\dd}_{\uu}&=&\chi^{\dd}_{0\uu}+\chi^{\uu}_{0\uu}V^{\uu}_{\uu}\chi^{\dd}_{\uu}+\chi^{\uu}_{0\uu}V^{\dd}_{\uu}\chi^{\dd}_{\dd}+\chi^{\dd}_{0\uu}V^{\uu}_{\dd}\chi^{\dd}_{\uu}+\chi^{\dd}_{0\uu}V^{\dd}_{\dd}\chi^{\dd}_{\dd} \\
\chi^{\uu}_{\dd}&=&\chi^{\uu}_{0\dd}+\chi^{\uu}_{0\dd}V^{\uu}_{\uu}\chi^{\uu}_{\uu}+\chi^{\uu}_{0\dd}V^{\dd}_{\uu}\chi^{\uu}_{\dd}+\chi^{\dd}_{0\dd}V^{\uu}_{\dd}\chi^{\uu}_{\uu}+\chi^{\dd}_{0\dd}V^{\dd}_{\dd}\chi^{\uu}_{\dd} \\
\chi^{\dd}_{\dd}&=&\chi^{\dd}_{0\dd}+\chi^{\uu}_{0\dd}V^{\uu}_{\uu}\chi^{\dd}_{\uu}+\chi^{\uu}_{0\dd}V^{\dd}_{\uu}\chi^{\dd}_{\dd}+\chi^{\dd}_{0\dd}V^{\uu}_{\dd}\chi^{\dd}_{\uu}+\chi^{\dd}_{0\dd}V^{\dd}_{\dd}\chi^{\dd}_{\dd}
\end{eqnarray*}
and
\begin{eqnarray*}
\chi^{\ud}_{\ud}&=&\chi^{\ud}_{0\ud}+\chi^{\ud}_{0\ud}V^{\ud}_{\ud}\chi^{\ud}_{\ud}+\chi^{\ud}_{0\ud}V^{\du}_{\ud}\chi^{\ud}_{\du}+\chi^{\du}_{0\ud}V^{\ud}_{\du}\chi^{\ud}_{\ud}+\chi^{\du}_{0\ud}V^{\du}_{\du}\chi^{\ud}_{\du} \\
\chi^{\du}_{\ud}&=&\chi^{\du}_{0\ud}+\chi^{\ud}_{0\ud}V^{\ud}_{\ud}\chi^{\du}_{\ud}+\chi^{\ud}_{0\ud}V^{\du}_{\ud}\chi^{\du}_{\du}+\chi^{\du}_{0\ud}V^{\ud}_{\du}\chi^{\du}_{\ud}+\chi^{\du}_{0\ud}V^{\du}_{\du}\chi^{\du}_{\du} \\
\chi^{\ud}_{\du}&=&\chi^{\ud}_{0\du}+\chi^{\ud}_{0\du}V^{\ud}_{\ud}\chi^{\ud}_{\ud}+\chi^{\ud}_{0\du}V^{\du}_{\ud}\chi^{\ud}_{\du}+\chi^{\du}_{0\du}V^{\ud}_{\du}\chi^{\ud}_{\ud}+\chi^{\du}_{0\du}V^{\du}_{\du}\chi^{\ud}_{\du} \\
\chi^{\du}_{\du}&=&\chi^{\du}_{0\du}+\chi^{\ud}_{0\du}V^{\ud}_{\ud}\chi^{\du}_{\ud}+\chi^{\ud}_{0\du}V^{\du}_{\ud}\chi^{\du}_{\du}+\chi^{\du}_{0\du}V^{\ud}_{\du}\chi^{\du}_{\ud}+\chi^{\du}_{0\du}V^{\du}_{\du}\chi^{\du}_{\du}
\end{eqnarray*}

To make it clear, we reorganize the above equations, and define:

\[
\bar{A}=
\begin{pmatrix}
 A^{\uu}_{\uu} & A^{\uu}_{\dd} \\
 A^{\dd}_{\uu} & A^{\dd}_{\dd}
\end{pmatrix}
\]

and

\[
\tilde{A}=
\begin{pmatrix}
 A^{\ud}_{\ud} & A^{\ud}_{\du} \\
 A^{\du}_{\ud} & A^{\du}_{\du}
\end{pmatrix}
\]

Then, the above equations become:
\begin{eqnarray*}
\bar{\chi}&=&\bar{\chi}_0+\bar{\chi}_0\bar{V}\bar{\chi} \\
\tilde{\chi}&=&\tilde{\chi_0}+\tilde{\chi_0}\tilde{V}\tilde{\chi}
\end{eqnarray*}

Note that the charge susceptibility ($\chi^C$) and longitudinal/transverse spin susceptibility ($\chi^{zz}$/$\chi^{\pm}$) will be:

\begin{eqnarray*}
\chi^{C}&=&\mathrm{Tr}[\chi^{\uu}_{\uu}+\chi^{\dd}_{\dd}+\chi^{\uu}_{\dd}+\chi^{\dd}_{\uu}] \\
\chi^{zz}&=&\mathrm{Tr}[\chi^{\uu}_{\uu}+\chi^{\dd}_{\dd}-\chi^{\uu}_{\dd}-\chi^{\dd}_{\uu}] \\
\chi^{+-}&=&\mathrm{Tr}[\chi^{\ud}_{\ud}] \\
\chi^{-+}&=&\mathrm{Tr}[\chi^{\du}_{\du}]
\end{eqnarray*}

\begin{figure}[h]
 \includegraphics[width=16cm]{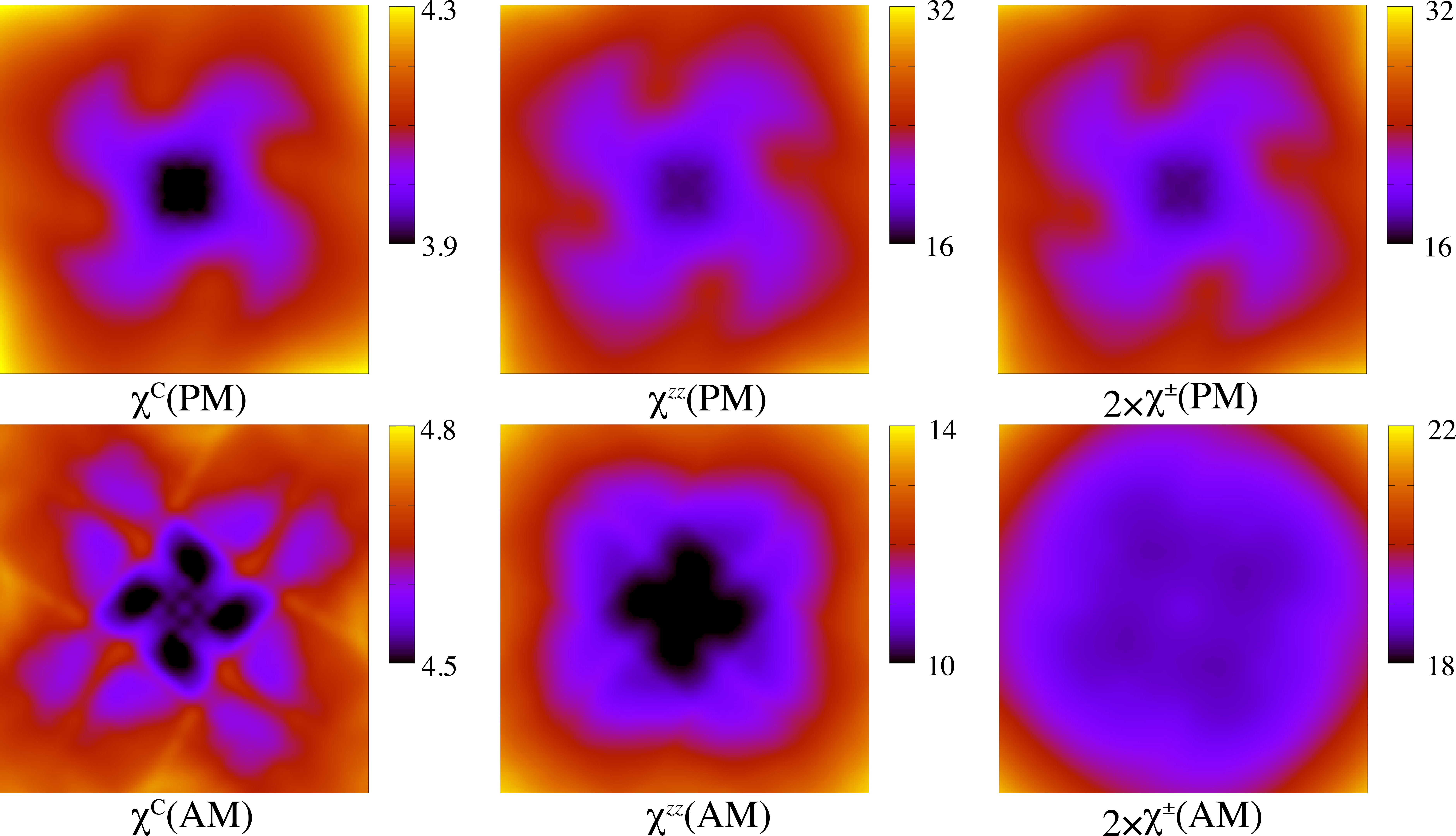}
\caption{The charge/spin susceptibility of \rb245\ in PM phase and \neelfm\ phase in $k_z=0$ plane. SU(2) symmetry ensures the identity $\chi^{zz}=2\chi^{\pm}$ in the PM case.\label{fig:chi}}
\end{figure}

The trace takes the form of $\mathrm{Tr}[\chi]=\sum_{s,p} \chi_{ss}^{pp}$. 

With these, the pairing interactions can now be calculated as:
$$[V(\mathbf{k}, \mathbf{k}')]_{tp\sigma_2\sigma_3}^{sq\sigma_1\sigma_4}=[U]_{st\sigma_1\sigma_2}^{pq\sigma_3\sigma_4}+[U\chi(\mathbf{k}'-\mathbf{k})U]_{st\sigma_1\sigma_2}^{pq\sigma_3\sigma_4}-[U\chi(\mathbf{k}'+\mathbf{k})U]_{sp\sigma_1\sigma_3}^{tq\sigma_2\sigma_4}$$

Since SOC is negligible and ignored in our calculations, we can define pairing channels using:
\begin{eqnarray*}
\Gamma^0(n'\mathbf{k}';n\mathbf{k})=&\sum  [V(\mathbf{k}, \mathbf{k}')]_{tp\sigma_2\sigma_3}^{sq\sigma_1\sigma_4} \times (\langle p\sigma_3 | \mathbf{k}n\uparrow\rangle \langle t\sigma_2 | -\mathbf{k}n\downarrow\rangle - \langle p\sigma_3 |\mathbf{k}n\downarrow\rangle \langle t\sigma_2 | -\mathbf{k}n\uparrow\rangle) \\
 &\times (\langle \mathbf{k}'n'\uparrow | q\sigma_4\rangle\langle -\mathbf{k}'n'\downarrow | s\sigma_1\rangle - \langle \mathbf{k}'n'\downarrow | q\sigma_4\rangle\langle -\mathbf{k}'n'\uparrow | s\sigma_1 \rangle)
\end{eqnarray*}

\begin{eqnarray*}
\Gamma^z(n'\mathbf{k}';n\mathbf{k})=&\sum  [V(\mathbf{k}, \mathbf{k}')]_{tp\sigma_2\sigma_3}^{sq\sigma_1\sigma_4} \times (\langle p\sigma_3 | \mathbf{k}n\uparrow\rangle \langle t\sigma_2 | -\mathbf{k}n\downarrow\rangle + \langle p\sigma_3 |\mathbf{k}n\downarrow\rangle \langle t\sigma_2 | -\mathbf{k}n\uparrow\rangle) \\
 &\times (\langle \mathbf{k}'n'\uparrow | q\sigma_4\rangle\langle -\mathbf{k}'n'\downarrow | s\sigma_1\rangle + \langle \mathbf{k}'n'\downarrow | q\sigma_4\rangle\langle -\mathbf{k}'n'\uparrow | s\sigma_1 \rangle)
\end{eqnarray*}

\begin{eqnarray*}
\Gamma^x(n'\mathbf{k}';n\mathbf{k})=&\sum  [V(\mathbf{k}, \mathbf{k}')]_{tp\sigma_2\sigma_3}^{sq\sigma_1\sigma_4} \times (\langle p\sigma_3 | \mathbf{k}n\uparrow\rangle \langle t\sigma_2 | -\mathbf{k}n\uparrow\rangle - \langle p\sigma_3 |\mathbf{k}n\downarrow\rangle \langle t\sigma_2 | -\mathbf{k}n\downarrow\rangle) \\
 &\times (\langle \mathbf{k}'n'\uparrow | q\sigma_4\rangle\langle -\mathbf{k}'n'\uparrow | s\sigma_1\rangle - \langle \mathbf{k}'n'\downarrow | q\sigma_4\rangle\langle -\mathbf{k}'n'\downarrow | s\sigma_1 \rangle)
\end{eqnarray*}

\begin{eqnarray*}
\Gamma^z(n'\mathbf{k}';n\mathbf{k})=&\sum  [V(\mathbf{k}, \mathbf{k}')]_{tp\sigma_2\sigma_3}^{sq\sigma_1\sigma_4} \times (\langle p\sigma_3 | \mathbf{k}n\uparrow\rangle \langle t\sigma_2 | -\mathbf{k}n\uparrow\rangle + \langle p\sigma_3 |\mathbf{k}n\downarrow\rangle \langle t\sigma_2 | -\mathbf{k}n\downarrow\rangle) \\
 &\times (\langle \mathbf{k}'n'\uparrow | q\sigma_4\rangle\langle -\mathbf{k}'n'\uparrow | s\sigma_1\rangle + \langle \mathbf{k}'n'\downarrow | q\sigma_4\rangle\langle -\mathbf{k}'n'\downarrow | s\sigma_1 \rangle)
\end{eqnarray*}

Notice that the summation over all orbital indices ($s,p,q,t$) and spin indicies ($\sigma_1,\sigma_2,\sigma_3,\sigma_4$) is understood. In addition, $\langle s\sigma' | \mathbf{k} n \sigma\rangle=\delta_{\sigma\sigma'}u_{\mathbf{k}n\sigma}^s$, where $ |\mathbf{k}n\sigma\rangle$ is the Bloch state with spin $\sigma$. The first channel $\Gamma^0$ is spin-singlet (SSP), the second channel $\Gamma^z$ is spin-triplet opposite spin pairing (OSP), and the rest two are spin-triplet equal-spin pairing (ESP).

Finally, the pairing strength $\lambda^l$ and gap function $\Delta^l(\mathbf{k})$ in channel $l=(0,x,y,z)$ can be solved using the linearized gap equation:

$$\lambda^l\Delta^l(\mathbf{k}')=-\frac{1}{V_{BZ}}\int_{FS}\frac{d^2k_{\parallel}}{\vert v_{\mathbf{k}}^{\perp}\vert} \Gamma^l(n'\mathbf{k'};n\mathbf{k})\Delta^l(\mathbf{k})$$

In the PM case, we also obtain a spin-less Hamiltonian by fitting first-principles results of nonmagnetic \rb245\ at 14 GPa using maximally projected Wannier function method. The non-interacting bare electron susceptibility is then calculated using:
$$[\chi^0(\mathbf{q}, i\nu)]_{st}^{pq}=-\frac{1}{N_{\mathbf{k}}\beta}\sum_{n,\mathbf{k}}G^0_{sp}(i\omega_n+i\nu, \mathbf{k+q})G^0_{qt}(i\omega_n, \mathbf{k})$$
Then we can construct the general bare susceptibility by setting:
$$[\chi^0(\mathbf{q}, i\nu)]_{st\sigma\sigma}^{pq\sigma\sigma}=[\chi^0(\mathbf{q}, i\nu)]_{st\sigma\bar{\sigma}}^{pq\sigma\bar{\sigma}}=[\chi^0(\mathbf{q}, i\nu)]_{st}^{pq}$$
and all the rest follows the same procedure.

With above method, we obtain the charge/spin susceptibilities in the PM and \neelfm\ states (FIG. \ref{fig:chi}), and obtained the pairing strengths presented in the maintext. All these calculations were performed with $32\times32\times32$ K-mesh with $\beta=640$ eV.

 \bibliography{245-p}